\documentclass[aps,prl,twocolumn,showpacs,superscriptaddress,floatfix]{revtex4-2}
\usepackage{graphicx}
\usepackage{amsmath}
\usepackage{amssymb}
\usepackage{color}
\usepackage{hyperref}

\begin{document}

\title{Interference of dynamical arrest, thermodynamic instabilities and energy-scale competition in symmetric binary mixtures}

\author{R. Peredo-Ortiz}
\affiliation{Instituto de F\'isica, Universidad Aut\'onoma de San Luis Potos\'i, \'Alvaro Obreg\'on 64, 78000 San Luis Potos\'i, M\'exico}

\author{E. L\'azaro-L\'azaro}
\affiliation{Instituto de F\'isica, Universidad Aut\'onoma de San Luis Potos\'i, \'Alvaro Obreg\'on 64, 78000 San Luis Potos\'i, M\'exico}

\author{M. Medina-Noyola}
\affiliation{Instituto de F\'isica, Universidad Aut\'onoma de San Luis Potos\'i, \'Alvaro Obreg\'on 64, 78000 San Luis Potos\'i, M\'exico}

\author{L.F. Elizondo-Aguilera}
\affiliation{Monte Caldera Technologies Inc., 12 N Cheyenne Ave, Tulsa, Oklahoma, USA}

\date{\today}

\begin{abstract}

While energy scale competition dictates equilibrium phase behavior, kinetic barriers often drive soft-matter mixtures into dynamically arrested states, rendering conventional phase diagrams incomplete. We extend the classification of binary systems into regions of thermodynamic instability to explore the interplay between phase separation and dynamical arrest. Strong cross-attractions kinetically suppress demixing, whereas competitive energy scales yield either condensation- or demixing-driven amorphous states. We show that this morphological crossover is parameterized by a structural order parameter, $\chi$, providing a unified non-equilibrium framework that reconciles theoretical predictions with experimentally observed arrested mixtures.

\end{abstract}

\maketitle

The formation of amorphous, dynamically arrested states in binary mixtures is driven by competing physical scales. For instance, in the paradigmatic case of size-asymmetric systems -- such as colloid-polymer, or binary hard-sphere suspensions -- the physical mechanisms underlying dynamical arrest depend strongly on the competition of length scales~\cite{Rigo2008, Edilio2018, Voigtmann_2011}. As the size-ratio $\delta = \sigma_{A}/\sigma_{B}$ decreases, the increasing disparity in geometric packing induces a bifurcation of the glass transition (GT), giving rise to the formation of different arrested states, such as repulsive (caging-driven) and attractive (depletion-driven) glasses, or double glasses in which the two species might arrest sequentially~\cite{Rosenfeld1989, Rigo2008, Roth2010, Voigtmann_2011, Edilio2018}. In these mixtures, thus, size-asymmetry determines the nature of the resulting amorphous states, explicitly governed by steric frustration between its components~\cite{Pusey1993, Poon1994, Poon1995, Dijkstra1999, Eckert2002, Ramakrishnan2002, Kozina2012}. 

In parallel, the description and classification of phase behavior in binary mixtures with attractive interactions is a foundational achievement of statistical thermodynamics. As first established by van Konynenburg and Scott~\cite{Konynenburg1980, Koefinger2006, Schoell2005} -- and later formalized by Chen and Forstmann~\cite{Bhatia, Forstmann} -- the topology of the phase diagram is mainly dictated by the energy-scale competition between the strength of self-species ($\epsilon_{ii}$) and cross-species ($\epsilon_{ij}$, $i\neq j$) attraction. Depending on the energy-ratio $\alpha = \epsilon_{ij}/\epsilon_{ii}$, the landscape of thermodynamic instability is fully determined: high $\alpha$ favors Gas-Liquid (GL) condensation, while low $\alpha$ favors Liquid-Liquid (LL) demixing. From a purely thermodynamic point of view, this discussion is well established: the instability temperatures -- the spinodal, describing GL condensation ($T_s$), and the $\lambda$-line, for LL demixing ($T_\lambda$) -- strictly delimit the boundaries of the equilibrium fluid phase~\cite{Pellicane2006, hansen, Koefinger2006, Koefinger2006_2, Schoell2005, Forstmann, JAMR_2005, Das2003, Das2006}.
 
For many soft matter mixtures, however, a thermodynamic phase diagram only represents an idealized picture of a more general non-equilibrium scenario. Experimental and simulated data~\cite{Lowen1997, Sciortino2005, Zaccarelli_2007, Gou2011, Varrato, Hsiao2015, Leheny}, for example, demonstrate that when such mixtures are quenched into their unstable regimes, the thermodynamic driving forces responsible for phase separation are often interrupted by dynamical arrest mechanisms. Consequently,  phase diagrams are insufficient to represent the experimental reality of these arrested systems, where the observed ``final states'' are the result of an intricate competition between instability and kinetic constraints~\cite{Edilio2018, Nohely_2023, JuanCarlos_2025}.

To push forward the understanding of these mixtures, thus, one requires a theoretical framework capable of extending their physical description into non-equilibrium regimes, particularly those involving instabilities and their interplay with dynamical arrest. To this end, in this contribution we employ the non-equilibrium self-consistent generalized Langevin equation (NESCGLE) theory~\cite{nescgle0, nescgle1, nescgle4, noneqOMtheory, review2026} (summarized in Sec. \textbf{S1} of the Supplementary Material (SM)) to analyze the irreversible evolution of model binary mixtures quenched towards different out-of-equilibrium domains. Unlike other theories of dynamical arrest, which usually treat thermodynamic instabilities only as singularities delimiting ``inaccessible'' regions in the control-parameter space~\cite{Evans1979, Goetze2009, scgle1}, NESCGLE also identifies these as boundary conditions for the loss of ergodicity, leading to a complex interplay between phase separation, dynamical arrest, and out-of-equilibrium relaxation~\cite{nescgle5, beny2021, gabysalr, Nohely_2023, JuanCarlos_2025}.

Specifically, we employ this framework to characterize symmetric binary mixtures (SBM, $\sigma_A=\sigma_B,\epsilon_{AA}=\epsilon_{BB}$) driven by competing attractive forces between species ($\epsilon_{ij} \neq \epsilon_{ii}$) and with variable energy ratio $\alpha$. In particular, we elucidate the resulting thermodynamic-instability and dynamical-arrest surfaces throughout the full concentration-density-temperature $(c, \rho,T)$ space (see  Secs. \textbf{S2} and \textbf{S3} of the SM). A central finding is the identification and classification of a variety of non-equilibrium amorphous states in SBM, emerging from the competition of GL condensation, LL demixing and different dynamical arrest transitions -- all of which are ultimately driven by $\alpha$. Thus, analogously to how size asymmetry induces different glassy-states~\cite{Voigtmann_2011, Rigo2008, Edilio2018}, here we show how energy competition prompts different arrested states in SBM.

In fact, we developed a ``Kinetic Atlas'' (detailed in Sec. \textbf{S3} of the SM) that comprehensively depicts the interplay between thermodynamic instabilities and dynamical arrest mechanisms across all relevant $\alpha$-values. This explicitly highlights how interference between the spinodal surface $T_s(\rho,c)$ (density-driven instability), the $\lambda$-surface $T_\lambda(\rho, c)$ (concentration-driven instability) and the dynamic arrest surface $T_c(\rho,c)$, is dictated by energy-scale competition. Importantly, our analysis of the asymptotic localization lengths also reveals that the GT surface $T_c$ does not end at its intersection with the spinodal surface, nor upon crossing the surface $T_\lambda(\rho,c)$.
Instead, it continues deep into the instability regions, thus extending to SBM the scenario previously identified for monodisperse systems ~\cite{nescgle5, beny2021, gabysalr}.

As detailed in Sec. \textbf{S4} of the SM (see Fig. S7), this extension of $T_c$ yields a ``gel-glass'' transition surface: a dynamic boundary separating the region of spatially-homogeneous ``attractive glasses'' from that of mesoscopically arrested ``gels'' formed via interrupted spinodal decomposition~\cite{nescgle5, nescgle7, nescgle8, beny2021}. As discussed in the following, this boundary also influences the kinetic evolution and final structure of a mixture after quenching inside the instability domains. Since the complete 3-dimensional topology of the phase/arrest  diagram of an SBM is intrinsically complex, here we focus -- without loss of generality -- on the equimolar case ($c=0.5$), to better highlight the most distinctive features introduced by energy competition. For concreteness, from now on we shall consider an SBM described by the hard-sphere plus square-well (HSSW) potential specified in Sec. \textbf{S2} of the SM.

\begin{figure}[t]
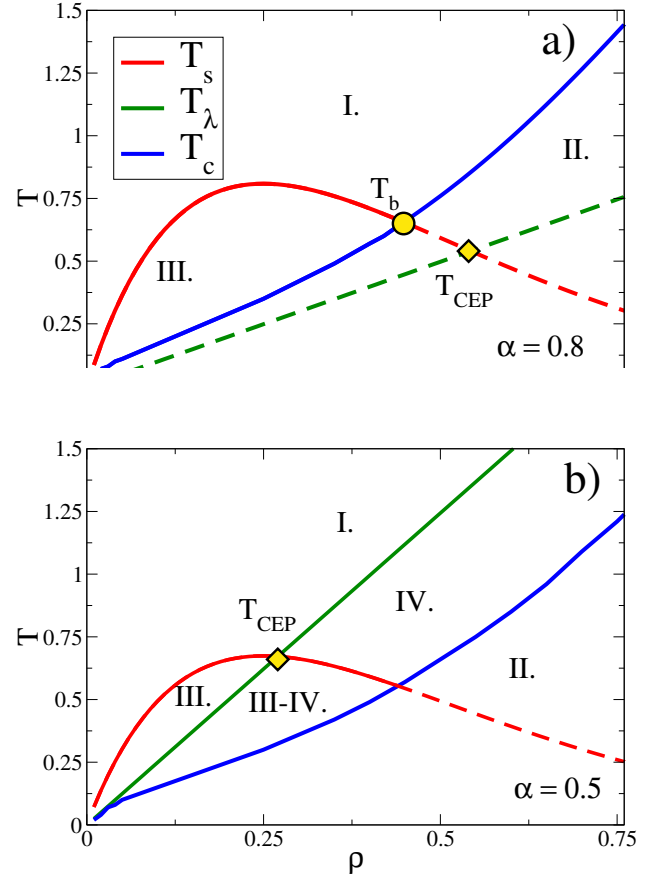

    \centering
    \includegraphics[width=0.95\linewidth]{Fig_1/alpha_0.8_c_0.5.eps}
    \includegraphics[width=0.95\linewidth]{Fig_1/alpha_0.5_c_0.5.eps}

    \caption{
    The competition between thermodynamic instability and dynamical arrest. 
    (a) \textbf{Demixing frustration} ($\alpha=0.8$): Strong cross-attraction causes the arrest line $T_c$ (blue) to detach from and remain strictly above the demixing instability (green). 
    The system vitrifies into a homogeneous glass, kinetically masking the demixing region. 
    (b) \textbf{Emergent Demixing} ($\alpha=0.5$): Weaker cross-attraction allows the demixing line to pierce the arrest line, creating a window for gelation. Inset: Locus of the bifurcation point; the intersection of the arrest line $T_c(\rho)$ with any instability occurs near the Hard-Sphere glass transition density.}
    \label{fig:suppression}
\end{figure}

For such a system, let us first discuss the case of relatively strong cross-attractions ($\alpha=0.8$, Fig. \ref{fig:suppression}a), where the strengths of self ($\epsilon_{ii}$) and cross-interactions ($\epsilon_{ij}$) are comparable, and thus the mixture is expected to display features of a monodisperse system ($\alpha=1$)~\cite{nescgle5, beny2021}. By analyzing the limit $k \to 0$ of elements $M_{\rho\rho}(k)$, $M_{cc}(k)$, and $M_{c\rho}(k)$ of the corresponding stability matrix $\mathbf{M}(k)$, the instability lines $T_s(\rho)$ and $T_\lambda(\rho)$ can be readily determined (see Subsection \textbf{S2.1} of the SM). These are illustrated here in Fig.~\ref{fig:suppression}a) by the red solid/dashed and green dashed lines, respectively. Notice that both curves intersect at the critical end point (CEP) $(\rho_{\text{\tiny{CEP}}},T_{\text{\tiny{CEP}}})$ (solid diamond), where LL demixing and GL condensation can coexist \cite{Koefinger2006}. Thus, this intersection defines two distinct density regimes for the occurrence of instability: (i) for $\rho<\rho_{\text{\tiny{CEP}}}$, $T_s>T_\lambda$, meaning that GL condensation precedes LL demixing upon cooling; and (ii), for $\rho>\rho_{\text{\tiny{CEP}}}$, $T_\lambda>T_s$, where LL demixing precedes GL condensation.

To complement the identification of thermodynamic instability regions in the $(\rho,T)$-plane, we also considered the non-ergodicity parameters (NEP) $\gamma_A^{(a)}$ and $\gamma_B^{(a)}$ of the NESCGLE, which allow to determine the dynamical arrest transition lines of the HSSW-SBM system (see Sec.~\textbf{S4} of the SM). From the analysis of these NEP over the full $(\rho,T)$-plane, we identify a discontinuous type B liquid-glass transition line $T_c(\rho)$ (blue solid line), which extends downward from its high-temperature high-density HS-like limit, and intersects the spinodal line at a bifurcation point $(\rho_b,T_b)$ (solid circle in Fig.\ref{fig:suppression}a). 


For $\rho\leq\rho_{\text{\tiny{b}}}(<\rho_{\text{\tiny{CEP}}})$, in addition, a bifurcation scenario for dynamical arrest is obtained (analogous to that found in the monodisperse case \cite{nescgle5, beny2021, review2026}). Specifically, a continuous (type A) dynamic arrest line is predicted to virtually coincide with $T_s(\rho)$ (superposed red solid line), indicating that a portion of the GL curve, besides being the threshold of thermodynamic instability, is also the threshold of non-ergodicity. Furthermore, the analysis of NEP for temperatures below $T_{s}$ also reveals the existence of another discontinuous (type B) transition that, just as in the monocomponent case, turns out to be a continuation of $T_c(\rho)$ inside the spinodal region. Remarkably, in terms of the NEP the demixing LL transition does not represent (nor interfere with) any kind of dynamical arrest condition. 

The resulting topology, which broadly depicts the interplay between different thermodynamic instabilities and dynamical arrest transitions, encompasses physical implications worth emphasizing, particularly in the regime $\rho<\rho_b$,  where $T_\lambda<T_c<T_s$. In this $\rho$-domain, NESCGLE predicts that the evolution of an equilibrium-liquid configuration (in region \textbf{I}) after a sudden temperature quench depends sensitively on the final temperature $T_f$. For a relatively shallow quench $T_c<T_f<T_s$ (region \textbf{III}), for example, the concomitant spinodal decomposition process that follows the quench is eventually interrupted by dynamical arrest conditions, so that the resulting arrested structure of the mixture retains memory of the interrupted GL separation (i.e., the structure factors show frozen spinodal decomposition peaks at low wave numbers $k$). For deeper quenches $T_f<T_c$, instead, dynamical arrest outpaces any thermodynamic instability, and the SBM vitrifies to a low-density, structurally homogeneous glassy state, completely suppressing the initial stages of spinodal decomposition (highlighted by the absence of the low-$k$ scattering peaks).

At higher densities ($\rho>\rho_b$), in contrast, a quench from equilibrium conditions to $T_f<T_c$ leads to the formation of glassy states with no signs of aggregation. Instead, the SBM exhibits liquid-like structures characterized by well-defined contact peaks at nearest-neighbor distances. Furthermore, in this high-density regime, neither spinodal decomposition nor segregation play any role in determining dynamical arrest conditions for the SBM. Remarkably, the above physical scenario is qualitatively consistent with both the experimental and simulated results for comparable binary mixtures~\cite{Varrato, Leheny, Foffi2005}.

For smaller energy ratios (e.g. $\alpha = 0.5$, Fig. \ref{fig:suppression}b), the attractive well associated with self-interactions ($\epsilon_{ii}$) is significantly deeper than that of cross-interactions ($\epsilon_{ij}$), thus favoring segregation of like-species within the mixture. 
In this case, the thermodynamic drive for demixing outweighs the influence of the glass transition, so that $T_\lambda$ rises strictly above $T_c$. Importantly, we found that the demixing line also describes a continuous type A transition, signaling the loss of ergodicity: slightly below $T_\lambda$ transient segregation precedes the dynamical arrest of the two species (see Subsection \textbf{S4.2} of the SM). Through its intersection with $T_s$ at the CEP, this emergent dynamical arrest condition creates a new landscape, partitioning the $(\rho,T)$-plane into distinct domains where spinodal decomposition actively competes with dynamical arrest.

In particular, a shallow quench into a region of instability  -- i.e. $T_f < \max(T_s, T_\lambda)$ --  induces a continuous non-ergodic transition. Depending on density, the quench is followed by either interrupted segregation ($\rho>\rho_\text{\tiny{CEP}}$), or arrested GL separation ($\rho<\rho_\text{\tiny{CEP}}$), leading to arrested states driven by different underlying mechanisms. In contrast, sufficiently deep quenches ($T_f<T_c$) always lead to discontinuous type B transitions, just as in the previous case (see Subsection \textbf{S4.2} of the SM).

The upward shift of $T_\lambda$ has additional implications that merit emphasis, particularly in the proximity of the CEP, where different physical mechanisms mingle. For densities slightly above $\rho_\text{\tiny{CEP}}$, for example, the analysis of the NEP suggests that only $T_\lambda$ represents a type A transition upon cooling, while $T_c$ defines a type B transition at lower temperatures. In this regime $T_s$ is simply unnoticed by the NEP. Conversely, for $\rho<\rho_\text{\tiny{CEP}}$ the line $T_s$ defines conditions for loss of ergodicity, whereas $T_\lambda$ is overlooked by the NEP which, for $T_f<T_s$, only identify $T_c$.  At first sight, this description -- restricted solely to the behavior of the NEP --  could misleadingly suggest that, once the type A ergodicity-breaking threshold is crossed (wether determined by $T_\lambda$ or $T_s$), the interplay between the two thermodynamic instabilities becomes largely irrelevant in determining the nature of the resulting arrested state. However, as we now discuss, this is not the case.

For clarity, let us highlight at his point the different regions of the parameter space $(\rho,T)$ in Fig. \ref{fig:suppression}b), delimited by $T_s$, $T_\lambda$ and $T_c$, which provide a generic classification of the distinct ergodic and arrested states predicted by the NESCGLE for the SBM. These regions correspond to: \textbf{I.} homogeneously-mixed (ergodic) fluid states, \textbf{II.} homogeneously arrested glassy states, \textbf{III.} gel-like states driven by arrested GL separation, and \textbf{IV.} partially demixed, bygel-like states. In addition, we identify a region denoted \textbf{III}-\textbf{IV}, representing a blurred crossover between glassy and instability-driven arrested states. Within this region, a diversity of amorphous structures emerges from an intricate interplay between thermodynamic instabilities and dynamical arrest.

To see this, we now turn to the discussion of the structural correlations predicted by NESCGLE for quenches inside region \textbf{III}-\textbf{IV}. In particular, we might consider the partial structure factors $ S_{ii}(k;t) \equiv N^{-1} \langle \delta \rho_i(\mathbf{k}, t) \delta \rho_i(-\mathbf{k},t)\rangle$. 
Here $\delta\rho_i(\mathbf{k}, t)\equiv  \langle\rho_i(\mathbf{k}, t)\rangle-\rho_i(\mathbf{k}, t)$, and $\rho_i(\mathbf{k};t)$ is the Fourier transform of the local density of the $i$-species, $\rho_i(t)=\sqrt{N_i}\sum_{l=1}^{N_i}\delta(\mathbf{r}-\mathbf{r}_l^{(i)}(t))$ \cite{PRE2025}. Although these canonical observables are routinely determined through scattering experiments (or simulations) to analyze structural behavior, under some circumstances they might display degenerate low wave-vector signals corresponding to stationary states driven by different underlying mechanisms~\cite{Das2003, Das2006}. For example, while colloid-polymer mixtures are driven by a demixing instability~\cite{Schmidt_2002, Shah2003}, the low-$k$ scattering signal obtained for dispersed colloids is qualitatively indistinguishable from that observed during condensation-driven cluster formation in attractive systems~\cite{wuliuchencao, Dorsaz2009, Gou2011}.

\begin{figure}[t]
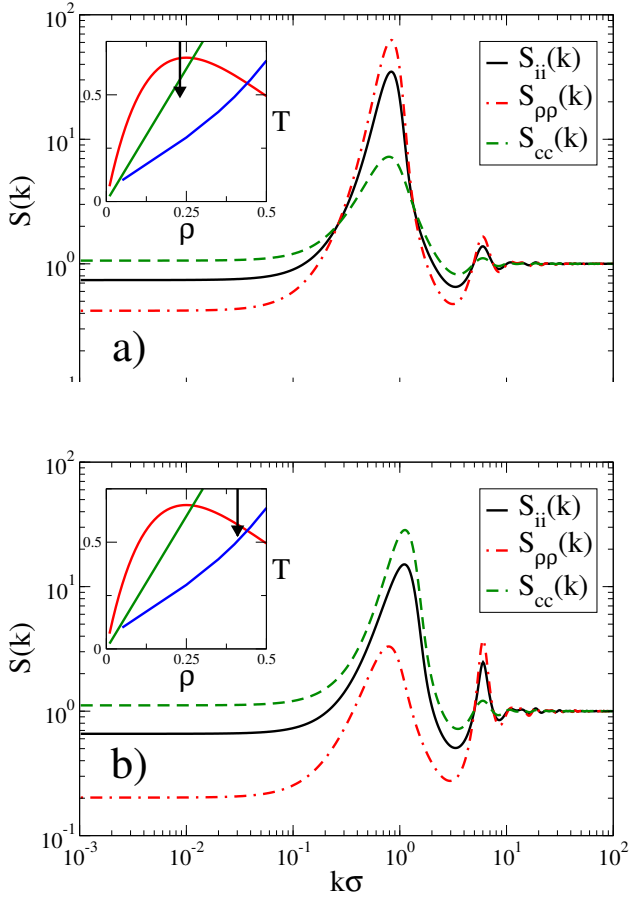

    \centering
    \includegraphics[width=0.95\linewidth]{Fig_2/new_rho_0.23_Tf_0.5.eps}
    \includegraphics[width=0.95\linewidth]{Fig_2/new_rho_0.41_Tf_0.5.eps}
    \caption{Avoidance of structural degeneracy via the Number-Concentration formalism ($\alpha = 0.5$). 
    (a) Close to the condensation regime, the low-$k$ peak is driven by density correlations $S_{\rho\rho}$ (dashed-dotted line, see the text). 
    (b) Close to the demixing regime, the IRO peak is driven by concentration correlations $S_{cc}$ (dashed line). Note that the standard partial structure factors $S_{ii}$ (solid lines) are qualitatively (and semiquantitatively) the same.}
    \label{fig:structure}
\end{figure}

To emphasize this notion, let us focus again on the case $\alpha=0.5$, and consider two isochoric quenches inside region \textbf{III}-\textbf{IV}, for densities below and above $\rho_\text{\tiny{CEP}}$. As illustrated in Figs. \ref{fig:structure}a ($\rho<\rho_\text{\tiny{CEP}}$) and Fig. \ref{fig:structure}b) ($\rho>\rho_\text{\tiny{CEP}}$), the asymptotic long-time structure factors $S^{(a)}_{ii}(k)\equiv \displaystyle{\lim_{t\to\infty}S_{ii}(k;t)}$ (solid lines) predicted by NESCGLE for these two quenches exhibit a prominent peak at low wavevectors, roughly $k_{IRO}\simeq 0.5$, which are indicative of mesoscopic Intermediate-Range-Order (IRO)~\cite{wuliuchencao, gabysalr}. Notice that, in both cases, these peaks are significantly higher than the corresponding nearest neighbors peaks, observed at $k_{main}\simeq 2\pi$. Hence, the two predicted structure factors $S_{ii}^{(a)}(k)$ display rather similar features, qualitatively and semi-quantitatively.

To elucidate whether these resulting structures are driven by arrested condensation or interrupted demixing processes, we now employ the number-concentration formalism~\cite{Bhatia, Forstmann}, which considers the transformation
\begin{align}
    S_{\rho\rho}^{(a)}(k) &= c_A S_{AA}^{(a)}(k) + c_B S_{BB}^{(a)}(k) + 2\sqrt{c_A c_B} S_{AB}^{(a)}(k) \label{eq:S_rhorho} \\
    S_{cc}^{(a)}(k) &= c_B S_{AA}^{(a)}(k) + c_A S_{BB}^{(a)}(k) - 2\sqrt{c_A c_B} S_{AB}^{(a)}(k) \label{eq:S_cc}
\end{align}
where $c_i=\rho_i/\rho$ ($i=A,B)$ are the molar fractions.


As illustrated in Fig. \ref{fig:structure}, this representation immediately highlights a reversal of roles for the correlation functions $S^{(a)}_{\rho\rho}(k)$ (dashed-dotted lines) and $S_{cc}^{(a)}(k)$ (dashed lines) predicted by NESCGLE. More specifically, for a quench closer to the condensation-driven regime \textbf{III} (Fig. \ref{fig:structure}a) the low-$k$ structure is dominated by the total density fluctuations $S_{\rho\rho}^{(a)}(k)$, which display a much larger IRO signal in comparison to that of $S_{cc}^{(a)}(k)$ (a difference of nearly one order of magnitude in the height of the IRO peak). For a quench closer to the demixing regime \textbf{IV} (Fig. \ref{fig:structure}b), instead, the IRO peak is dominated by concentration correlations $S_{cc}^{(a)}$, whereas $S^{(a)}_{ii}(k)$ and $S_{\rho\rho}^{(a)}(k)$ capture the influence of the nearest-neighbor signal.



\begin{figure}[t]
    \centering
    \includegraphics[width=0.95\linewidth]{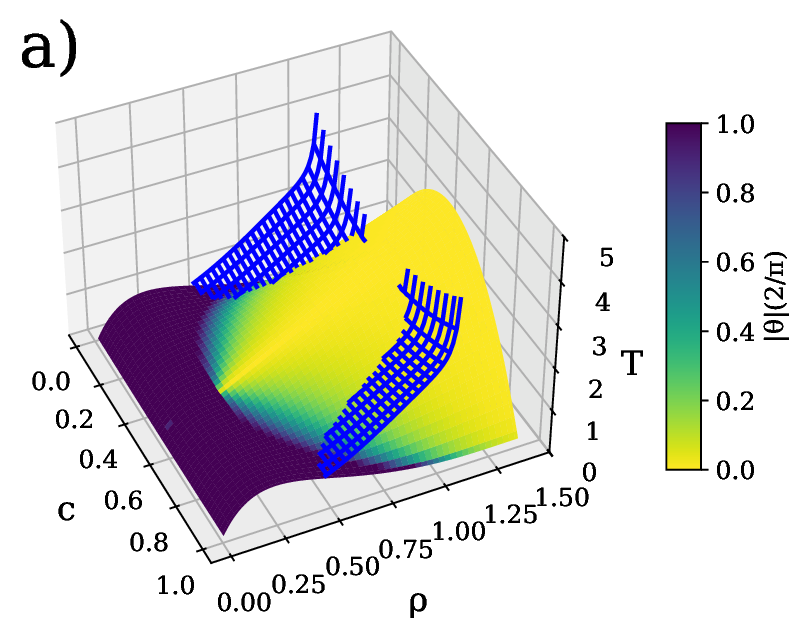}
    \includegraphics[width=0.95\linewidth]{Fig_3/alpha_0.5_c_0.5_chi.eps}
    \caption{From Thermodynamic Instability to Non-Equilibrium Arrest for SBM with $\alpha=0.5$. (a) \textbf{Thermodynamic Prediction:} The 3D instability surface colored by the equilibrium instability angle $|\theta|$ (Dark: Condensation, Light: Demixing). The blue wireframe indicates the dynamic arrest surface $T_c$. Thermodynamics predicts the driving force only at the instability limit. (b) \textbf{Non-Equilibrium Reality:} A cross-section of the Kinetic Atlas at $c=0.5$, classified by the structural order parameter $\chi$. Here, the energy scale $\alpha$ is tuned to allow demixing. The parameter $\chi$ acts as the non-equilibrium extension of $\theta$, classifying the state deep into the arrested region. The color gradient visualizes the continuous morphological transition from density-driven gels ($\chi\to 1$) to concentration-driven gels ($\chi\to 0$).}
    \label{fig:chi_map}
\end{figure}

To further rationalize the dynamical arrest landscape of an SBM, in Fig. \ref{fig:chi_map}a) we now consider the spinodal decomposition surface in the full parameter space $(\rho,T,c)$, for the case $\alpha=0.5$. This surface, however, is described using the so-called instability angle $|\theta|$ -- defined in \cite{Forstmann}, and in section \textbf{S3} of SM -- and represented here using the color grade pallet on the right side of the figure. As explained in Ref. \cite{Forstmann}, this quantity describes the system's propensity at the onset of instability: darker regions (i.e. $|\theta|\to \pi/2$) indicate a drive toward condensation, while lighter regions ($|\theta|\to 0$) indicate a drive toward demixing. For completeness, Fig. \ref{fig:chi_map}a) also considers the dynamical arrest surface $T_c(\rho,c)$ (blue mesh) which, as discussed previously, penetrates the spinodal surface defining different arrested regimes. Let us highlight here that a (thermodynamic) description based solely in $|\theta|$ is fundamentally restricted to the charcterization of the instability surface, and cannot provide any insight of the final structure of a mixture quenched deep into the non-equilibrium regions discussed previously.


Hence, to depict the structural characteristics of an SBM quenched inside the instability regimes, we might consider a non-equilibrium counterpart of the instability angle, namely, a weighting parameter $\chi$, defined here as:
\begin{equation}
    \chi = \frac{S^{(a)}_{\rho\rho}(k_{IRO})}{S^{(a)}_{\rho\rho}(k_{IRO}) + S^{(a)}_{cc}(k_{IRO})},
\end{equation}
which quantifies the influence of $S^{(a)}_{\rho\rho}$ and $S^{(a)}_{cc}$ on the resulting arrested structure factor $S^{(a)}_{ii}(k)$. Fig. \ref{fig:chi_map}b) illustrates this notion, by considering a cut (at $c=0.5$) of the 3D surface in Fig. \ref{fig:chi_map}a). Thus represented, one sees that $\chi$ describes a smooth color gradient inside Region \textbf{III}-\textbf{IV}, starting from a darker domain [$\chi\to1, S_{cc}(k_{IRO})<<S_{\rho\rho}(k_{IRO})$] that corresponds to arrested states influenced by GL separation; and evolving gradually into an increasingly lighter domain [$\chi\to0, S_{\rho\rho}(k_{IRO})<<S_{cc}(k_{IRO})$], describing frozen structures influenced by LL demixing. This color gradient, thus, provides a more vivid representation of the competition of instabilities and dynamical arrest at the level of $S_{ii}^{(a)}(k)$, allowing to classify different amorphous states in SBMs.

Evaluated on the instability surface, the parameter $\chi$ completely recovers the information provided by $\theta$ and, hence, can be regarded as a non-equilibrium extension of the instability angle $\theta$, and can be employed to classify amorphous states of SBMs quenched inside instability regions. Furthermore, as shown in Section \textbf{S3} of the SM, the parameter $\chi$ can be also used to systematically evaluate the influence of energy-scale competition (i.e. $\alpha)$ on the arrested states of an SBM, generated by the interference of different thermodynamical instability and dynamical arrest processes. To the best of our knowledge, this is the first time these aspects have been addressed from a theoretical perspective. 

In summary, we have provided a brief theoretical characterization of the behavior of symmetric binary mixtures, quenched into different  non-equilibrium conditions, involving the simultaneous interference of thermodynamical instabilities and dynamical arrest mechanisms. Specifically, we analyzed the influence of energy-scale competition, which turns out to play a similar role to that of length-scale competition in purely repulsive glass-forming systems. Analogously to how size-asymmetry produces different glassy states in hard-sphere binary mixtures, a significant energy disparity generates a rich diversity of dynamically arrested states, that derive from the interplay of two instabilities -- gas-liquid separation and liquid-liquid demixing -- and two (type A and type B) transitions describing ergodicity-loss. 


Our methodology, based on the capability of the parameter $\chi$ to identify and classify the nature of arrested states in mixtures, opens clear pathways for future research. For example, the present work focused only on the analysis of asymptotic structural properties. The full NESCGLE framework, however, allows for the characterization of out-of-equilirbium dynamical and viscoelastic properties  \cite{JoR2025}, such as the storage $G'(\omega; t)$ and loss $G''(\omega; t)$ moduli, allowing us to directly correlate the microstructural fingerprints identified here with specific rheological signatures, accessible to conventional experimental methods. Furthermore, in this work we assumed --  only for simplicity -- instantaneous quenching protocols. As demonstrated recently, however, there are no fundamental barriers to include more realistic processing conditions \cite{PoF2025}, such as cooling/heating, or compression rates, which strongly influence the properties of non-equilibrium matter \cite{Royal2012, Hsiao2015, Scott2023}.


\begin{acknowledgments}
This work was supported by Secretar\'ia de Ciencia, Humanidades, Tecnolog\'ia e Innovaci\'on (SECIHTI) through Postdoctoral Fellowships Grants No. I1200/224/2021 and I1200/320/2022. 
This work was also partially supported by Monte Caldera Technologies.
\end{acknowledgments}

\bibliographystyle{unsrt}
\bibliography{biblio}

\end{document}


\maketitle

\setcounter{section}{0}
\renewcommand{\thesection}{S\arabic{section}}

\setcounter{figure}{0}
\renewcommand{\thefigure}{S\arabic{figure}}

\numberwithin{equation}{section}
\renewcommand{\theequation}{\thesection.\arabic{equation}}

\section{The NESCGLE theory}

This supplemental material summarizes the fundamental concepts underlying the Non-Equilibrium Self-Consistent Generalized Langevin Equation (NESCGLE) theory \cite{nescgle1, nescgle4, noneqOMtheory, review2026}. 
We restrict our description to the governing equations for the waiting-time dependent structure factor, defined as $S_{ij}(k;t)\equiv(1/N)\angles{\delta\rho_i(k,t)\delta\rho_j(-k,t)}^{(ss)}$.
The notation $\langle\dots\rangle^{ss}$ denotes an average over a stationary state, generalizing the standard thermodynamic equilibrium ensemble average \cite{PRE2025}.
Under thermodynamic equilibrium conditions, this structure factor is intimately related to the Stability Matrix, $\mathcal{E}_{ij}$, defined as:
\begin{equation}\label{eq:SM_C}
    \mathcal{E}_{ij}(k;\rho_A, \rho_B, T) \equiv \delta_{ij} - \sqrt{\rho_i\rho_j}c_{ij}(k),
\end{equation}
where $c_{ij}(k)$ is the direct correlation function obtained from the Ornstein-Zernike Equation (OZE), once a closure relation is specified~\cite{hansen, mcquarrie}; and the diagonal density matrix $\sqrt{\bm{\rho}}$ has elements defined as $\sqrt{\rho}_{ij}\equiv \delta_{ij}\sqrt{\rho_i}$.

In its simplest form, the NESCGLE theory provides a framework to compute the structural relaxation of a colloidal system following a sudden change in its thermodynamic state. 
The time evolution of the non-equilibrium structure factor, $\bm{S}(k;t)$ is governed by the dynamic law:
\begin{equation}\label{eq:S(k;t)}
    \begin{split}
        \frac{\partial \bm{S}(k;t)}{\partial t} = & \bm{H}(t) \cdot \braces{\bm{S}(k;t)-\brakets{\sqrt{\bm{\rho}^{(f)}}\cdot\bm{\mathcal{E}}\cdot \sqrt{\bm{\rho}^{(f)}}}^{-1}}\\
        & + \braces{\bm{S}(k;t)-\brakets{\sqrt{\bm{\rho}^{(f)}}\cdot\bm{\mathcal{E}}\cdot \sqrt{\bm{\rho}^{(f)}}}^{-1}}\cdot\bm{H}^\dagger(t),
    \end{split}
\end{equation}
here the $2\times 2$ relaxation rate matrix is defined as $\bm{H}(t) = -k^2\bm{D}^0\cdot\bm{b}(t)\cdot\brakets{\sqrt{\bm{\rho}^{(f)}}\cdot\bm{\mathcal{E}}\cdot \sqrt{\bm{\rho}^{(f)}}}$.
The diagonal matrices $\bm{D}^0$ and $\bm{b}(t)$ represents the short-time self-diffusion coefficients and the time-dependent mobility functions of the particles, respectively.

The elements of the matrix $[\bm{b}(t)]_{ij} = \delta_{ij}b_{ii}(t)$ serve as the state parameters governing the dynamical system defined by Eq. \eqref{eq:S(k;t)}. 
The elements of the diagonal time-dependent mobility functions are evaluated as:
\begin{equation}
    b_{ii}(t) = \brakets{1 + \int_0^\infty d\tau\Delta\zeta_{ii}(\tau;t)}^{-1}.
\end{equation}
The friction memory functions, $\Delta\zeta_{ii}$ are obtained by solving the \textit{Self-Consistent Scheme} (SCS), defined by the following system of equations:
\begin{equation}\label{eq:DeltaZeta}
    \begin{split}
        \Delta\zeta_{ii}(\tau;t) = & \frac{D_i^0}{3(2\pi)^3}\int d^3k k^2 \brakets{\bm{F}^{(s)}(k,\tau;t)}_{ii}\left[\bm{h}(k;t)\cdot\sqrt{\bm{\rho}}\cdot\bm{S}^{-1}(k;t)\right.\\
        & \left.\cdot \bm{F}(k,\tau;t)\cdot\bm{S}^{-1}(k;t)\cdot\sqrt{\bm{\rho}}\cdot\bm{h}(k;t)\right]_{ii},
    \end{split}
\end{equation}
where the time-dependent total correlation matrix is given by $\bm{h}(k;t) = \sqrt{\bm{\rho}^{-1}}\cdot\pbrakets{\bm{S}(k;t)-\bm{I}}\cdot\sqrt{\bm{\rho}^{-1}}$.
In Laplace space, the self- and collective intermediate scattering functions, $\bm{F}^{(s)}(k,\tau;t)$ and $\bm{F}(k,\tau;t)$, are given respectively by
\begin{equation}\label{eq:Fs(k,z;t)}
    \begin{split}
        \bm{F}^{(s)}(k,z;t) = \braces{z\bm{I} + k^2\bm{D}^0\cdot\brakets{\bm{I}+\bm{\Lambda}(k)\cdot\bm{\Delta\zeta}(z)}^{-1}}^{-1},
    \end{split}
\end{equation}
and
\begin{equation}\label{eq:F(k,z;t)}
    \begin{split}
        \bm{F}(k,z;t) = & \left\lbrace z\bm{I} + k^2\bm{D}^0\cdot\brakets{\bm{I}+\bm{\Lambda}(k)\cdot\bm{\Delta\zeta}(z;t)}^{-1}
        \cdot\bm{S}^{-1}(k;t)\right\rbrace^{-1}\cdot\bm{S}^{-1}(k;t).
    \end{split}
\end{equation}
Here, the elements of the matrix $\bm{\Lambda}(k)$ are defined as $\Lambda_{ij}(k)=\delta_{ij}[1+(k/k_i^{(c)})]$, where the empirical cut-off wave-vector is $k_i^{(c)} = 1.305(2\pi)/\sigma_i$.
Within the NESCGLE, this parameter is employed to \textit{calibrate} the theory for each specific application\cite{mendoza}.

The Symmetric Binary Mixtures (SBM) offers the advantage of identical particle sizes, implying that the short-time self-diffusion coefficients are identical, $D_1^0=D^0_2=D^0$.
Furthermore, at the equimolar concentration, $c=0.5$, symmetry dictates that there is no physical distinction between the dynamics of the two species, leading to identical mobility functions, $b(t)=b_1(t)= b_2(t)$.
However, dynamic asymmetry (one species arresting before the other) is excluded by symmetry in this specific study, but is a feature of the general theory.
Consequently, the problem can be assigned to the same methodologies as previously defined for monodisperse systems \cite{nescgle3}.

Under these conditions, for an instantaneous quench from an initial temperature $T_i$ ($t<0$) to a final temperature $T$ ($t>0$), Eq. \eqref{eq:S(k;t)} admits an analytic solution of the form
\begin{equation}\label{eq:S(k;u)}
    \begin{split}   
        \bm{S}(k;u) =& \brakets{\sqrt{\bm{\rho}_f}\cdot\bm{\mathcal{E}}^{(f)}\cdot \sqrt{\bm{\rho}_f}}^{-1} + \exp \left[-\bm{A}(k,u)\right]\\
        & \cdot\braces{\bm{S}(k;t=0)-\brakets{\sqrt{\bm{\rho}_f}\cdot\bm{\mathcal{E}}^{(f)}\cdot \sqrt{\bm{\rho}_f}}^{-1} }\cdot \exp \left[-\bm{A}^\dagger(k,u)\right].
    \end{split}
\end{equation}
The decay matrix is defined as $\bm{A}(k,u) = uk^2\bm{D}^0\cdot[\sqrt{\rho_f}\cdot \bm{\mathcal{E}}^{(f)}(k)\cdot \sqrt{\rho_f}]$, where the stability matrix is a function of the final thermodynamic state $\bm{\mathcal{E}}^{(f)} = \bm{\mathcal{E}}(\rho_f, T_f)$.
The material time, or \textit{internal time},  $u(t)$ represents the accumulated mobility of the system~\cite{InnerClocks} and is defined as
\begin{equation}\label{eq:u(t)}
    u(t) = \int_0^t dt' b(t').
\end{equation}
Due to the symmetry at $c=0.5$, the material times for both species are identical, $u(t)=u_1(t)=u_2(t)$, reflecting the symmetry of their respective mobilities, $b_{ii}(t)=b(t)$.

A practical approach to studying the non-equilibrium properties of a system is to evaluate the asymptotic structure factor, defined as $\bm{S}^{(a)}(k)\equiv\bm{S}(k;t\to\infty)$.
Under equilibrium conditions—specifically for temperatures situated above the condensation, demixing, and ideal glass transition lines—the system remains ergodic. 
Consequently, the material time diverges at long times $\lim_{t\to\infty}u(t)$.
As a result, the elements of the decay matrix $\bm{A}(k,u)$ diverge, causing the exponential relaxation terms in Eq. \eqref{eq:S(k;u)} to vanish. 
Thus, the asymptotic structure factor recovers the standard equilibrium form $\bm{S}^{(a)}(k) = [\sqrt{\rho_f}\cdot \bm{\mathcal{E}}^{(f)}(k)\cdot \sqrt{\rho_f}]^{-1} = \bm{S}^{eq}(k)$.

However, crossing any of the transition lines causes the asymptotic structure factor, $\bm{S}^{(a)}(k)$, to deviate from the equilibrium form $[\sqrt{\bm{\rho}_f}\cdot \bm{\mathcal{E}}^{(f)}(k)\cdot \sqrt{\bm{\rho}_f}]^{-1}$.
In this regime, $\bm{S}^{(a)}(k)$ becomes dependent on the specific preparation protocol used to reach the final thermodynamic state $(\rho_f, T_f)$ \cite{nescgle1, noneqOMtheory, review2026}.
To determine the asymptotic structure $\bm{S}^{(a)}(k)$, we must evaluate the localization lengths $\gamma_i$ for species $i=A, B$, defined as the long-time asymptotic limits of the mean squared displacements, $\gamma_i \equiv \lim_{\tau\to\infty} \langle |\Delta \bm{r}_i(\tau)|^2 \rangle$.

These lengths are obtained by solving the following system of coupled bifurcation equations \cite{Rigo2008}:
\begin{equation}\label{eq:gammas}
    \begin{split}
        \frac{1}{\gamma_i} = &  \frac{1}{3(2\pi)^3}\int d^3k k^2 \left[ \bm{\Lambda} (\bm{\Lambda} + k^2\bm{\gamma})^{-1} \right]_{ii} \\
        & \times\left[ \bm{c}(t)\sqrt{\bm{\rho}} \bm{S}(t) \bm{\Lambda} (\bm{S}(t)\bm{\Lambda} + k^2\bm{\gamma}^{(a)})^{-1} \sqrt{\bm{\rho}} \bm{h}(t)\right]_{ii},    
    \end{split}
\end{equation}
which are obtained in the asymptotic limit $\tau\to\infty$ of the SCS described by equations \eqref{eq:DeltaZeta}-\eqref{eq:F(k,z;t)}.

In our calculation, we replace the static structure factor in Eq. \eqref{eq:gammas} with the material-time dependent structure factor $\bm{S}(k;u)$ defined in Eq. \eqref{eq:S(k;u)}.
Consequently, the localization length becomes a function of the material time, $\bm{\gamma} = \bm{\gamma}(u)$.
Under equilibrium conditions, $\bm{\gamma}(u)$ diverges for all $u$. 
However, under non-equilibrium conditions, there exists a finite value of $u$ where the localization length transitions to a finite value\cite{InnerClocks}.
This limiting value defines the asymptotic material time, $u^{(a)} \equiv \lim_{t\to\infty} u(t)$.
As a result, the asymptotic localization length is given by $\bm{\gamma}^{(a)} = \bm{\gamma}(u^{(a)})$.

The solutions to Eq. \eqref{eq:gammas} reveal two distinct dynamical regimes: an \textit{ergodic region} where $1/\gamma_i^{(a)} = 0$ (infinite localization length, fluid state), and a \textit{non-ergodic region} where $1/\gamma_i^{(a)} > 0$ (finite localization length, arrested state).
The boundary between these regions defines the \textit{critical arrest temperature}, $T_c(\rho, c)$.

\section{Model System: The Symmetric Binary Mixture}

This supplemental material provides methodological details related to the model of Symmetric Binary Mixture (SBM).
We consider a binary mixture composed of $N_A$ particles of species $A$ and $N_B$ particles of species $B$ contained in a volume $V$. 
The total number density is given by $\rho = (N_A+N_B)/V$, and the molar fractions are defined as $c_i = N_i/(N_A+N_B)$ for $i=A,B$. 
In SBM, the effective diameters of both species are equal, $\sigma_A = \sigma_B = \sigma$.

The thermodynamic behavior of this kind of mixture is determined by the pairwise interaction potential $U_{ij}(r)$. In this work, we specifically model the SBM using a Hard-Sphere plus Square-Well (HS-SW) potential, in which  the interaction between species $i$ and $j$ is defined as:
\begin{equation}\label{eq:HSSW}
    U_{ij}(r) =
        \begin{cases}
            \infty & \text{for } r < \sigma_{ij}, \\
            -\epsilon_{ij} & \text{for } \sigma_{ij} \le r < \lambda^{(SW)}_{ij}, \\
            0 & \text{for } r \ge \lambda^{(SW)}_{ij}.
        \end{cases}
\end{equation}
where $\sigma_{AA}=\sigma_{BB}=\sigma_{AB}=\sigma_{BA}=\sigma$. The energy scales are defined such that $\epsilon_{AA}=\epsilon_{BB}=\epsilon$ (self-attraction) and $\epsilon_{AB}=\alpha\epsilon$ (cross-attraction). 
The parameter $\lambda^{(SW)}_{ij}$ defines the range of the attractive well. 
Hence, The parameter $\alpha$ determines the competition between condensation and demixing.


Two distinct regimes emerge based on the value of $\alpha$. 
In the first case, $\alpha > 1$, the cross-interaction is stronger than the self-interactions, and the system exhibits only Gas-Liquid (GL) condensation \cite{Forstmann}. 
The second case, $\alpha < 1$, presents a much richer phenomenology characterized by the competition between GL condensation and Liquid-Liquid (LL) demixing as a function of the thermodynamic state variables $c = c_A$ and $\rho$ \cite{Forstmann, Schoell2005, Koefinger2006, Leheny, Varrato}.


\subsection{Thermodynamic Stability}

We generalize the approach of Sharma and Sharma~\cite{Sharma} to the case of binary mixtures. 
In their original work, the authors approximated the direct correlation function of a monodisperse colloidal system interacting via a hard-sphere plus square-well potential as the sum of a reference hard-sphere term and a perturbative attractive tail: $c(r) \approx c^{HS}(r) - \beta u^{SW}(r)$~\cite{Sharma}. 
Here, $c^{HS}(r)$ is the hard-sphere direct correlation function~\cite{Baxter} and $u^{SW}_{ij}(r)$ is the attractive square-well potential.
In thermodynamic equilibrium, the structure factor is determined entirely by this matrix via the relation \eqref{eq:SM_C}. 

To determine the boundaries of thermodynamic stability (where the equilibrium relation $\bm{S}(k) = [\sqrt{\bm{\rho}}\cdot\bm{\mathcal{E}}(k)\cdot\sqrt{\bm{\rho}}]^{-1}$ is not satisfied), we analyze the eigenvalues of the stability matrix. 
Following Chen and Forstmann \cite{Forstmann}, the eigenvalues $\lambda_{1,2}(k)$ are given by:
\begin{equation}\label{eq:lambda_12}
    \lambda_{1,2}(k) = \frac{\mathcal{E}_{AA}(k) + \mathcal{E}_{BB}(k) \mp \sqrt{\left[ \mathcal{E}_{AA}(k) - \mathcal{E}_{BB}(k) \right]^2 + 4 \mathcal{E}_{AB}^2(k)}}{2}.
\end{equation}
The instability surface, marking the limit of stability, is defined by the condition where the smaller eigenvalue vanishes in the long-wavelength limit: $\lambda_1(k \to 0) = 0$.

A key advantage of the formalism developed by Chen and Forstmann \cite{Forstmann} is the ability to transform these correlation functions from the particle-species basis into the collective basis of total number density ($\rho = \rho_1+\rho_2$) and concentration ($c = \rho_1/\rho$) fluctuations. 
The transformed matrix $\bm{M}(k)$ has elements defined as:
\begin{align}
    M_{\rho\rho}(k) &= c_1 \mathcal{E}_{11}(k) + c_2 \mathcal{E}_{22}(k) + 2\sqrt{c_1 c_2} \mathcal{E}_{12}(k), \label{eq:M_rho_rho_A} \\
    M_{cc}(k) &= c_1c_2 \left[ c_2 \mathcal{E}_{11}(k) + c_1 \mathcal{E}_{22}(k) - 2\sqrt{c_1 c_2} \mathcal{E}_{12}(k) \right], \label{eq:M_cc_A} \\
    M_{\rho c}(k) &= c_1c_2 \left[ \mathcal{E}_{11}(k) - \mathcal{E}_{22}(k) \right] + \sqrt{c_1 c_2}(c_2 - c_1) \mathcal{E}_{12}(k). \label{eq:M_rho_c_A}
\end{align}

In the thermodynamic limit ($k \to 0$), these elements are directly related to the second derivatives of the Helmholtz and Gibbs free energy.
Specifically, $M_{\rho\rho}$ relates to the Helmholtz free energy $A$ via the isothermal compressibility:
\begin{equation}\label{eq:Helmholtz}
    \left.\frac{\partial^2 (A/V)}{\partial\rho^2}\right|_{T,c} = \frac{k_BT}{\rho}M_{\rho\rho}(0),
\end{equation}
and $M_{cc}$ relates to the Gibbs free energy $G$ via the osmotic stability:
\begin{equation}\label{eq:Gibbs}
    \left( \frac{\partial^2 G}{\partial c^2} \right)_{T,P,N} = N k_B T \left[ M_{cc}(0) - \frac{M_{\rho c}(0)^2}{M_{\rho\rho}(0)} \right].
\end{equation}
To quantify which mechanism drives the instability (GL condensation vs. LL demixing), we compute the instability angle $\theta$:
\begin{equation}\label{eq:theta}
    \tan(\theta) = \frac{2 M_{\rho c}(0)}{M_{\rho\rho}(0) - M_{cc}(0)}.
\end{equation}
If $|\theta| \approx 0$, the instability is predominantly LL demixing; if $|\theta| \approx \pi/2$, it is predominantly GL condensation. 
The conditions where Eq. \eqref{eq:Helmholtz} is zero defines de Spinodal Surface $T_s(\rho, c)$, and equivalently the conditions to make Eq. \eqref{eq:Gibbs} zero defines the $\lambda$-Surface $T_\lambda(\rho,c)$.
Intermediate values indicate a coupled instability\cite{Varrato}.

\section{Kinetic Atlas: Topological Regimes}

This supplement provides the complete three-dimensional (3D) visualizations of the thermodynamic instability surfaces, $T_\text{Inst}(\rho, c)$, and the dynamic arrest surfaces, $T_c(\rho, c)$, for the Symmetrical Binary Mixture (SBM). 
These plots complement the 2D projections presented in the main text and offer a direct view of the topological changes that occur as the cross-species interaction parameter $\alpha$ is varied.

The classification of these topologies follows the rigorous scheme for binary fluids established by van Konynenburg and Scott~\cite{Konynenburg1980}, and adapted specifically for symmetrical mixtures by Sch\"oll-Paschinger et al.~\cite{Schoell2005} and K\"ofinger et al.~\cite{Koefinger2006}. 
By varying $\alpha$, we traverse the global phase diagram, identifying five distinct regimes:
\begin{itemize}
    \item \textbf{Type I-A ($\alpha > 1$):} Pure condensation (demixing is thermodynamically impossible)~\cite{Fantoni2005}.
    \item \textbf{Type I ($\alpha \approx 0.8$):} Demixing frustration (demixing exists but is buried).
    \item \textbf{Type II ($\alpha \approx 0.6$):} Emergent demixing (arrest intersects the demixing ridge).
    \item \textbf{Type III ($\alpha \approx 0.3$):} Tricritical merger (coalescence of instability lines).
    \item \textbf{Type IV ($\alpha \approx 0.0$):} Pure demixing (condensation vanishes)~\cite{Schoell2005}.
\end{itemize}

For each topological type, we present two viewing angles: a ``Front View'' showing the general structure in $(c, \rho, T)$ space, and a ``Back View'' to highlight features obscured by the primary surface curvature.

The instability surface is defined by the condition $\lambda_1(c, \rho, T) = 0$ (where $\lambda_1$ is the smallest eigenvalue of the stability matrix), which marks the onset of either gas-liquid (GL) condensation or liquid-liquid (LL) demixing~\cite{Forstmann, Fantoni2005}. 
The dynamic arrest surface is defined by the divergence of the relaxation time within the Self-Consistent Generalized Langevin Equation (SCGLE) theory \cite{Rigo2008, Rigo2008_2, Edilio2018, Nohely_2023, JuanCarlos_2025}.

\subsection{Type I-A Topology ($\alpha > 1$): Pure Condensation}

\begin{figure}[H]
    \centering
    \begin{subfigure}[b]{0.48\textwidth}
        \centering
        \includegraphics[width=1.0\linewidth]{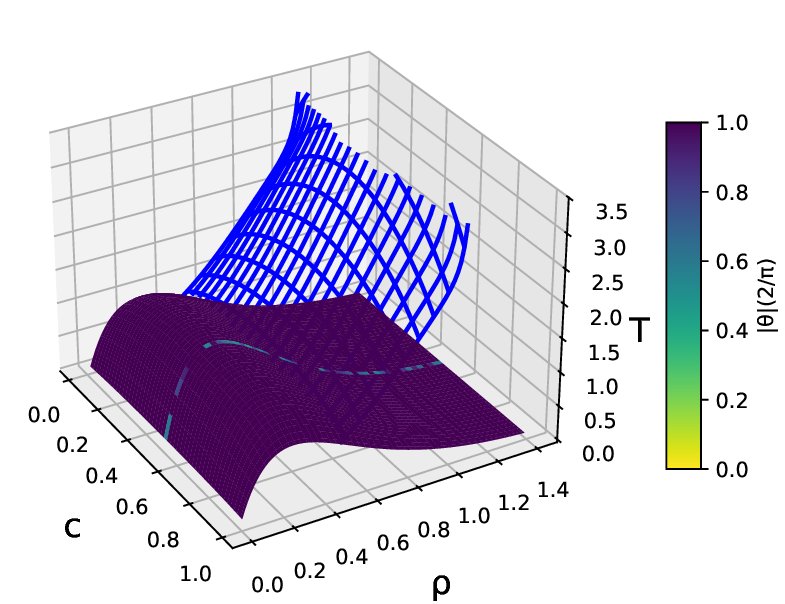}
        \caption{Front View}
    \end{subfigure}
    \hfill
    \begin{subfigure}[b]{0.48\textwidth}
        \centering
        \includegraphics[width=1.0\linewidth]{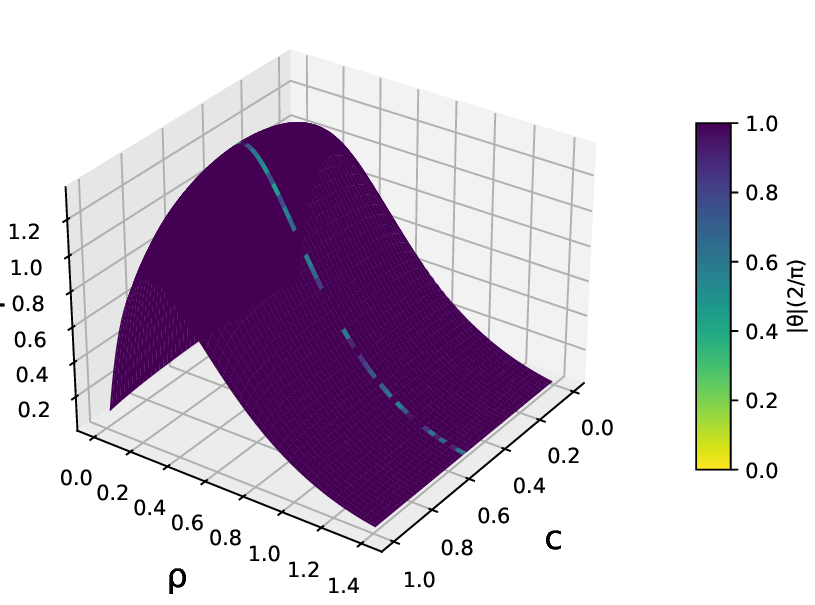}
        \caption{Back View}
    \end{subfigure}
    \caption{
    Type I-A Topology ($\alpha=2.0$): Pure Condensation.
    The 3D instability surface $T_\text{inst}(\rho, c)$ manifests as a single, symmetric convex dome centered at equimolar concentration. 
    The surface coloration, mapped to the instability angle $|\theta|$, is uniformly dark, indicating that the instability is driven exclusively by total density fluctuations (Gas-Liquid condensation). 
    This topology confirms the analytical prediction~\cite{Fantoni2005} that strong cross-attraction ($\alpha > 1$) renders demixing thermodynamically impossible.
    The dynamic arrest surface $T_c$ (blue mesh) intersects this condensation dome, predicting a direct transition from fluid to a homogeneous attractive glass via arrested spinodal decomposition.}
    \label{fig:S1}
\end{figure}

Figure \ref{fig:S1} illustrates the regime where cross-species attraction dominates ($\alpha=2.0$), classified as Type I-A in the generalized van Konynenburg and Scott scheme~\cite{Koefinger2006}. 
Thermodynamically, the excess strength of the cross-interaction ($\epsilon_{ij} > \epsilon_{ii}$) creates a strong energetic preference for mixing. 
As demonstrated analytically by Fantoni et al.~\cite{Fantoni2005}, the condition for a demixing spinodal cannot be met in this regime because the relative adhesiveness is too high; consequently, the liquid-liquid demixing instability is thermodynamically suppressed. 
The resulting instability surface $T_\text{Inst}(\rho, c)$ manifests as a single, symmetric \textit{convex dome} centered at the equimolar concentration, corresponding exclusively to Gas-Liquid (GL) condensation driven by total density fluctuations.

Kinetically, the dynamic arrest surface (represented by the blue wireframe) intersects this condensation dome, reproducing the topology characteristic of \textit{arrested spinodal decomposition} observed in simple attractive fluids~\cite{beny2021}. 
This intersection implies that deep quenches into the unstable region lead to gelation arrested by local caging and bonding, rather than phase separation. 
This topology confirms that in the limit of strong cross-attraction, the binary nature of the system is effectively masked. 
The solidification is driven solely by density fluctuations ($S_{\rho\rho}$) without competition from concentration fluctuations ($S_{cc}$), reducing the phenomenological behavior to that of an effective single-component fluid.

\subsection{Type I ($\alpha \approx 0.8$): Demixing frustration}

As the cross-species interaction strength decreases to $\alpha = 0.8$, the system transitions into the Type I topological regime, following the classification of Sch\"oll-Paschinger et al.~\cite{Schoell2005} and K\"ofinger et al.~\cite{Koefinger2006}. 
Thermodynamically, the instability landscape remains dominated by a prominent Gas-Liquid (GL) condensation dome. 
However, unlike the Type I-A regime, a latent Liquid-Liquid (LL) demixing instability now exists. 
This instability surface is located at low temperatures, buried deep within the high-density region of the phase diagram. 
The intersection of the demixing line with the condensation binodal defines a Critical End Point (CEP) which, in this regime, lies significantly below the liquid-vapor critical temperature ($T_{CEP} < T_{cr}$)~\cite{JAMR_2005}.

\begin{figure}[H]
    \centering
    \begin{subfigure}[b]{0.48\textwidth}
        \centering
        \includegraphics[width=1.0\linewidth]{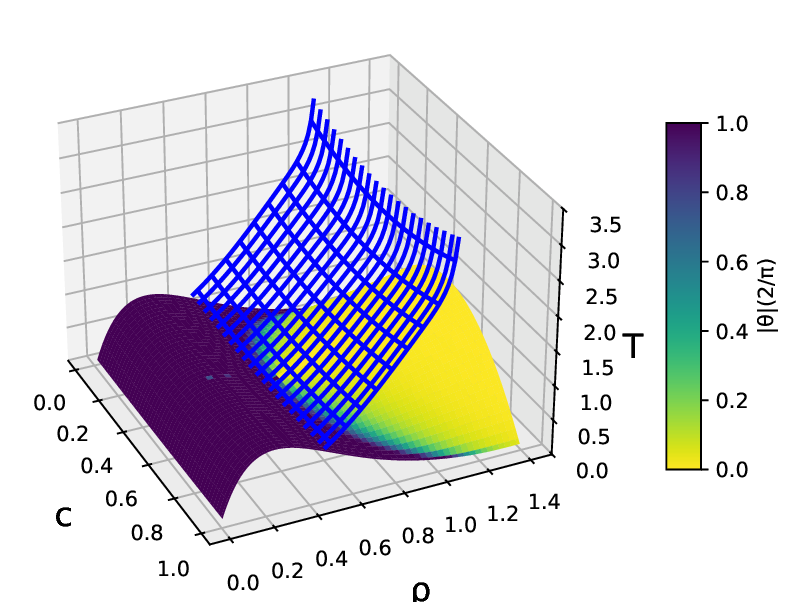}
        \caption{Front View}
    \end{subfigure}
    \hfill
    \begin{subfigure}[b]{0.48\textwidth}
        \centering
        \includegraphics[width=1.0\linewidth]{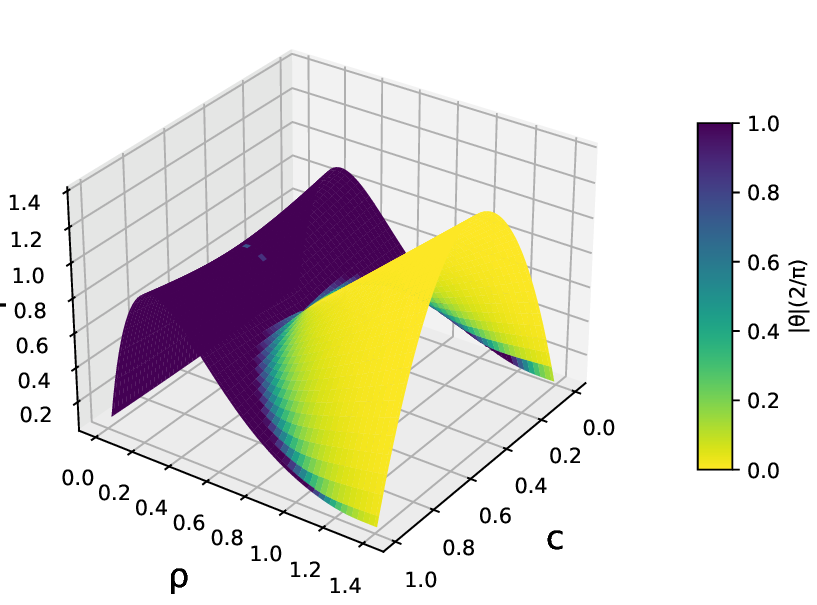}
        \caption{Back View}
    \end{subfigure}
    \caption{
    Type I Topology ($\alpha=0.8$): Thermodynamic frustration. 
    The 3D thermodynamic instability surface is dominated by the dark Gas-Liquid (GL) condensation dome. 
    Consistent with the Type I classification \cite{Koefinger2006}, a latent Liquid-Liquid (LL) demixing instability (lighter region) exists but is buried deep at low temperatures, with a Critical End Point (CEP) located significantly below the liquid-vapor critical temperature ($T_{CEP} < T_{cr}$). 
    Crucially, the dynamic arrest surface $T_c$ (blue mesh) lies strictly above this latent demixing region. 
    This topology visualizes the phenomenon of \textit{Thermodynamic frustration}: the glass transition acts as a kinetic shield, trapping the system in a homogeneous \textit{double glass} state strictly before it can access the thermodynamic driving force for phase separation.}
    \label{fig:S2}
\end{figure}

Kinetically, this regime is characterized by the phenomenon of \textit{Demixing frustration}. 
The dynamic arrest surface, $T_c(\rho, c)$ (represented by the blue wireframe in Fig. \ref{fig:S2}), is positioned strictly above the nascent demixing instability. 
This topological arrangement has profound physical consequences: for any isochoric quench from the fluid phase, the system inevitably intersects the glass transition line before it can access the thermodynamic demixing instability. 
Consequently, the slow dynamics effectively ``hides'' the thermodynamic drive for phase separation. 
The arrest surface acts as a kinetic shield, trapping the system in a homogeneous \textit{double glass} strictly before the thermodynamic preference for demixing can manifest.

\subsection{Type II ($\alpha \approx 0.6$): Emergent demixing}

\begin{figure}[H]
    \centering
    \begin{subfigure}[b]{0.48\textwidth}
        \centering
        \includegraphics[width=1.0\linewidth]{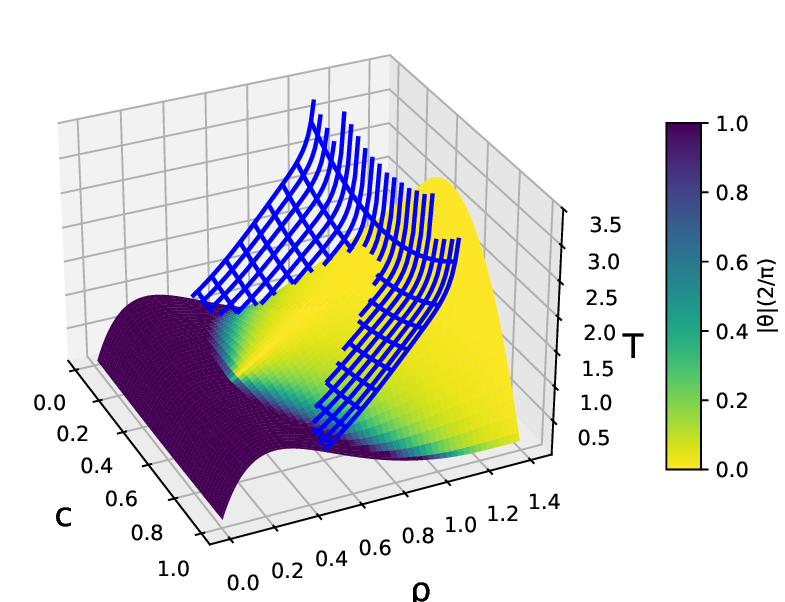}
        \caption{Front View}
    \end{subfigure}
    \hfill
    \begin{subfigure}[b]{0.48\textwidth}
        \centering
        \includegraphics[width=1.0\linewidth]{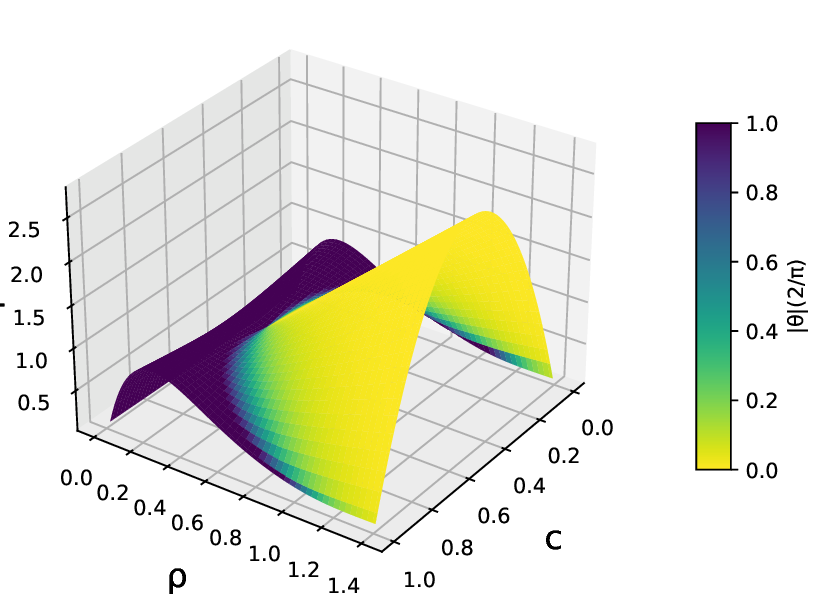}
        \caption{Back View}
    \end{subfigure}
    \caption{
    Type II Topology ($\alpha=0.6$): Emergent Demixing. 
    As the cross-species attraction weakens, the system transitions to the Type II regime~\cite{Koefinger2006}. 
    Thermodynamically, the liquid-liquid demixing instability rises in temperature, piercing the gas-liquid condensation dome. 
    This is visualized by the emergence of a light-colored ridge (indicating concentration-driven instability, $|\theta| \approx 0$) disrupting the smooth, dark condensation surface (density-driven, $|\theta| \approx \pi/2$). 
    The intersection of these surfaces marks the Critical End Point (CEP)~\cite{JAMR_2005}. 
    Kinetically, the arrest surface $T_c$ (blue mesh) now intersects this exposed demixing ridge. 
    This topology creates a \textbf{kinetic bifurcation}: unlike Type I, high-density quenches now access the demixing instability before arrest, opening a pathway for gelation driven by arrested phase separation.}
    \label{fig:S3}
\end{figure}

Figure \ref{fig:S3} captures the transition to the Type II topological regime ($\alpha=0.6$), corresponding to the classification established by K\"ofinger et al.~\cite{Koefinger2006}. 
In this regime, the thermodynamic landscape undergoes a critical topological shift: as the self-attraction becomes comparable to the cross-attraction, the previously buried liquid-liquid (LL) demixing instability rises significantly in temperature, eventually piercing the gas-liquid (GL) condensation dome. 
This emergence manifests visually as a distinct ``demixing ridge'' (lighter color) that disrupts the smooth convex curvature of the condensation surface. 
The precise intersection of the GL and LL instability surfaces defines a line of Critical End Points (CEP), which shifts toward higher temperatures as $\alpha$ decreases~\cite{JAMR_2005}.

Kinetically, this regime marks the \textit{breakdown of suppression} and the onset of a \textit{Kinetic Bifurcation}~\cite{nescgle5, nescgle7, nescgle8, beny2021, gabysalr}. 
Unlike the Type I case where the glass transition completely shields the demixing instability, the dynamic arrest surface $T_c$ (blue wireframe) now intersects the emerging demixing ridge at high densities. 
This topology creates a fork in the system's kinetic fate determined by the quench density: a quench at low density encounters the condensation dome first, leading to a gel formed by arrested spinodal decomposition; conversely, a quench at high density intersects the exposed demixing ridge first, driving the system toward arrested phase separation. 
This bifurcation confirms that the emergence of the demixing instability creates a new window for gelation driven by concentration fluctuations ($S_{cc}$), distinct from the density-driven arrest ($S_{\rho\rho}$) of the monodisperse limit.

\subsection{Type III ($\alpha \approx 0.3$): Tricritical merger} 

\begin{figure}[H]
    \centering
    \begin{subfigure}[b]{0.48\textwidth}
        \centering
        \includegraphics[width=1.0\linewidth]{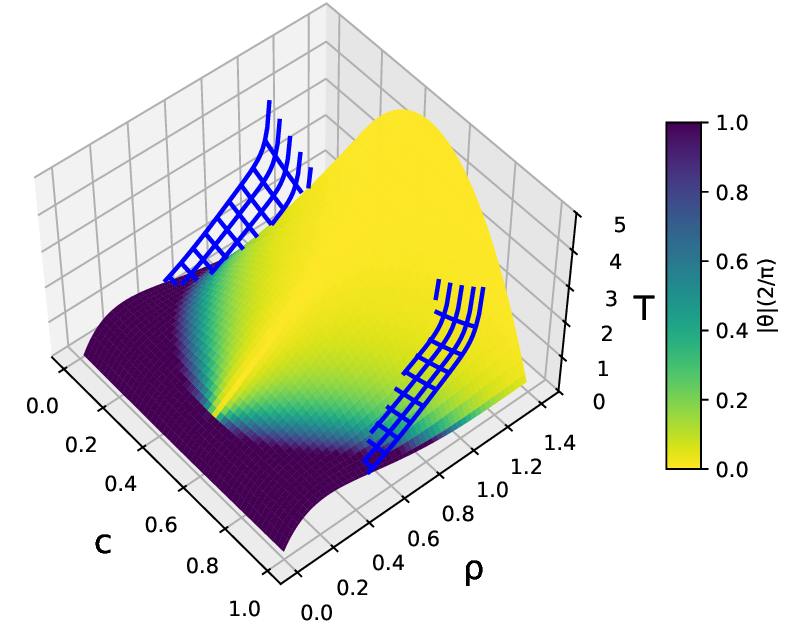}
        \caption{Front View}
    \end{subfigure}
    \hfill
    \begin{subfigure}[b]{0.48\textwidth}
        \centering
        \includegraphics[width=1.0\linewidth]{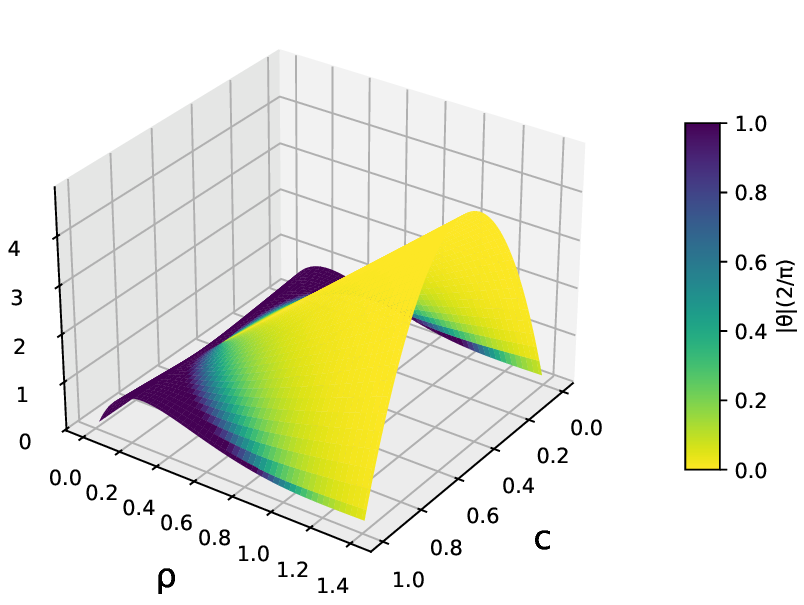}
        \caption{Back View}
    \end{subfigure}
    \caption{
    Type III Topology ($\alpha=0.3$): Tricritical Merger. 
    In this regime, the Gas-Liquid and Liquid-Liquid instability surfaces merge into a single continuous sheet. 
    The point of coalescence defines the \textit{Tricritical Point (TCP)}, where the distinction between condensation and demixing vanishes. 
    Kinetically, the arrest surface (blue mesh) flattens significantly compared to Type I/II. 
    The disappearance of the deep conditions to obtain $T_c$ indicates that the cross-species bonding—which stabilizes the attractive glass in higher $\alpha$ regimes—is no longer sufficient to generate a distinct re-entrant glass transition, leading to a smooth crossover between density-arrested and concentration-arrested states.}
    \label{fig:S4}
\end{figure}

Figure \ref{fig:S4} illustrates the Type III regime ($\alpha=0.3$), defined by the coalescence of instability mechanisms. 
As the cross-species interaction strength drops further, the miscibility gap expands significantly, leading to a fundamental topological change described by Sch\"oll-Paschinger et al.~\cite{Schoell2005}: the gas-liquid (GL) and liquid-liquid (LL) instability lines merge at a Tricritical Point (TCP). 
Unlike the Type II regime where the demixing ridge pierces the condensation dome, here the two surfaces fuse into a continuous sheet~\cite{JAMR_2004}. 
Thermodynamically, this implies that at the TCP, the three coexisting phases (vapor, A-rich liquid, and B-rich liquid) become identical simultaneously. 
Across this continuous surface, the instability mechanism transitions smoothly from density-driven at the flanks ($|\theta| \approx \pi/2$) to concentration-driven at the center ($|\theta| \approx 0$).

Kinetically, this regime is marked by a significant flattening of the dynamic arrest surface $T_c$. 
The characteristic deepness in the arrest temperature—prominent in Types I and II due to strong cross-species bonding—diminishes. 
This reflects the weakening of the cross-species attraction ($\epsilon_{ij}$); as $\alpha$ decreases, the driving force for gelation shifts from local attractive bonding (caging) toward global phase separation, resulting in a kinetic surface that smoothly bridges the density-driven and concentration-driven arrest boundaries.

\subsection{Type IV ($\alpha \approx 0.0$): Pure demixing }

\begin{figure}[H]
    \centering
    \begin{subfigure}[b]{0.48\textwidth}
        \centering
        \includegraphics[width=1.0\linewidth]{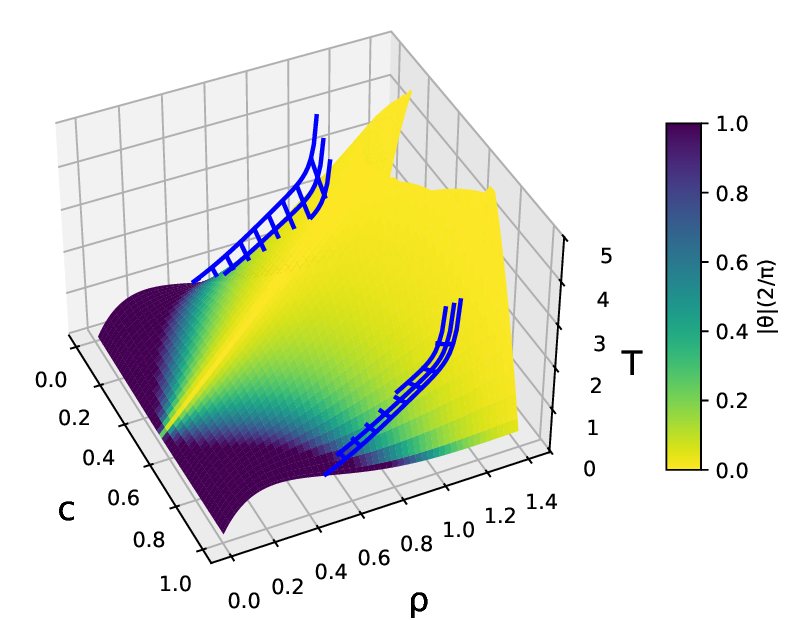}
        \caption{Front View}
    \end{subfigure}
    \hfill
    \begin{subfigure}[b]{0.48\textwidth}
        \centering
        \includegraphics[width=1.0\linewidth]{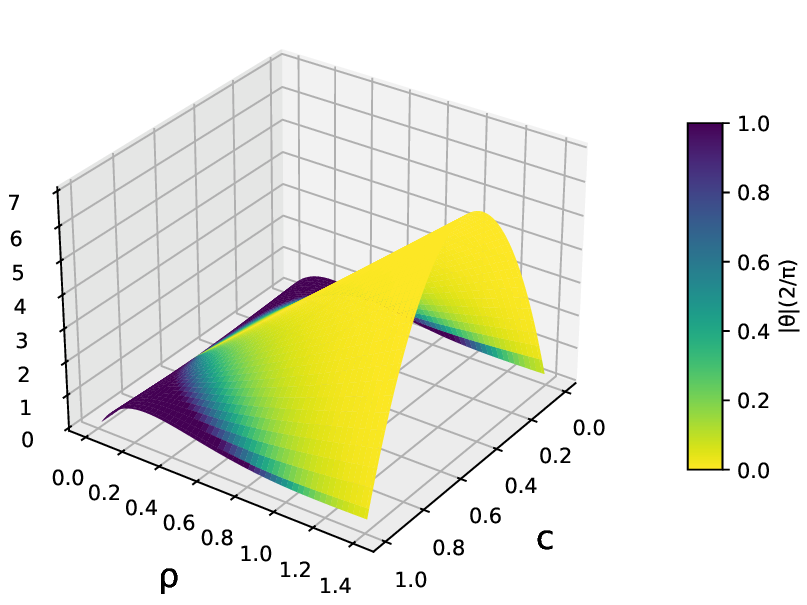}
        \caption{Back View}
    \end{subfigure}
    \caption{
    Type IV Topology ($\alpha=0.0$): Pure Demixing. 
    In the limit of zero cross-species attraction, the thermodynamic landscape is fundamentally altered: the condensation dome vanishes, replaced by vertical sheets indicating a pure demixing instability ($S_{cc}$) that extends to low densities. 
    Kinetically, the arrest surface (blue mesh) decouples from the temperature-dependent attraction and becomes a vertical wall located at the hard-sphere glass transition density. 
    This topology confirms that without cross-species attraction to form a stabilizing network (attractive glass), the system cannot arrest at low densities and instead undergoes macroscopic phase separation until the high-density repulsive glass limit is reached.}
    \label{fig:S5}
\end{figure}

Figure \ref{fig:S5} presents the Type IV limit ($\alpha=0.0$), representing the extreme case of non-interacting species (in terms of attraction), as described by Sch\"oll-Paschinger et al.~\cite{Schoell2005}. 
Thermodynamically, the absence of cross-species attraction ($\epsilon_{ij} \to 0$) renders mixing purely entropic and highly unfavorable compared to the energetic gain of self-association. 
Consequently, the convex Gas-Liquid condensation dome vanishes entirely. 
It is replaced by vertical instability sheets that extend toward zero density, indicating that the system is unstable to phase separation at nearly all temperatures and densities. 
In this regime, the instability is driven exclusively by concentration fluctuations ($S_{cc}$), corresponding to a mechanism of Pure Demixing~\cite{Fantoni2005}.

Kinetically, this regime exhibits a dramatic Kinetic Decoupling. 
While thermodynamics predicts demixing instability across a vast region of the phase diagram, the dynamic arrest surface transforms into a \textit{vertical wall} fixed at the Hard-Sphere glass transition density ($\phi_g \approx 0.58$)~\cite{nescgle3, Pusey1986}. 
Unlike the Type I or II regimes, where strong cross-attraction ($\epsilon_{ij}$) facilitates gelation at low densities via an attractive glass mechanism, the absence of cross-linking in the Type IV limit prevents low-density arrest. 
The system faces a binary fate: it either phase separates completely via fluid-fluid demixing or, at sufficiently high total packing fractions, jams into a repulsive glass driven solely by excluded volume interactions.

\subsection{Iso-Temperature Contour Maps}

\begin{figure}[H]
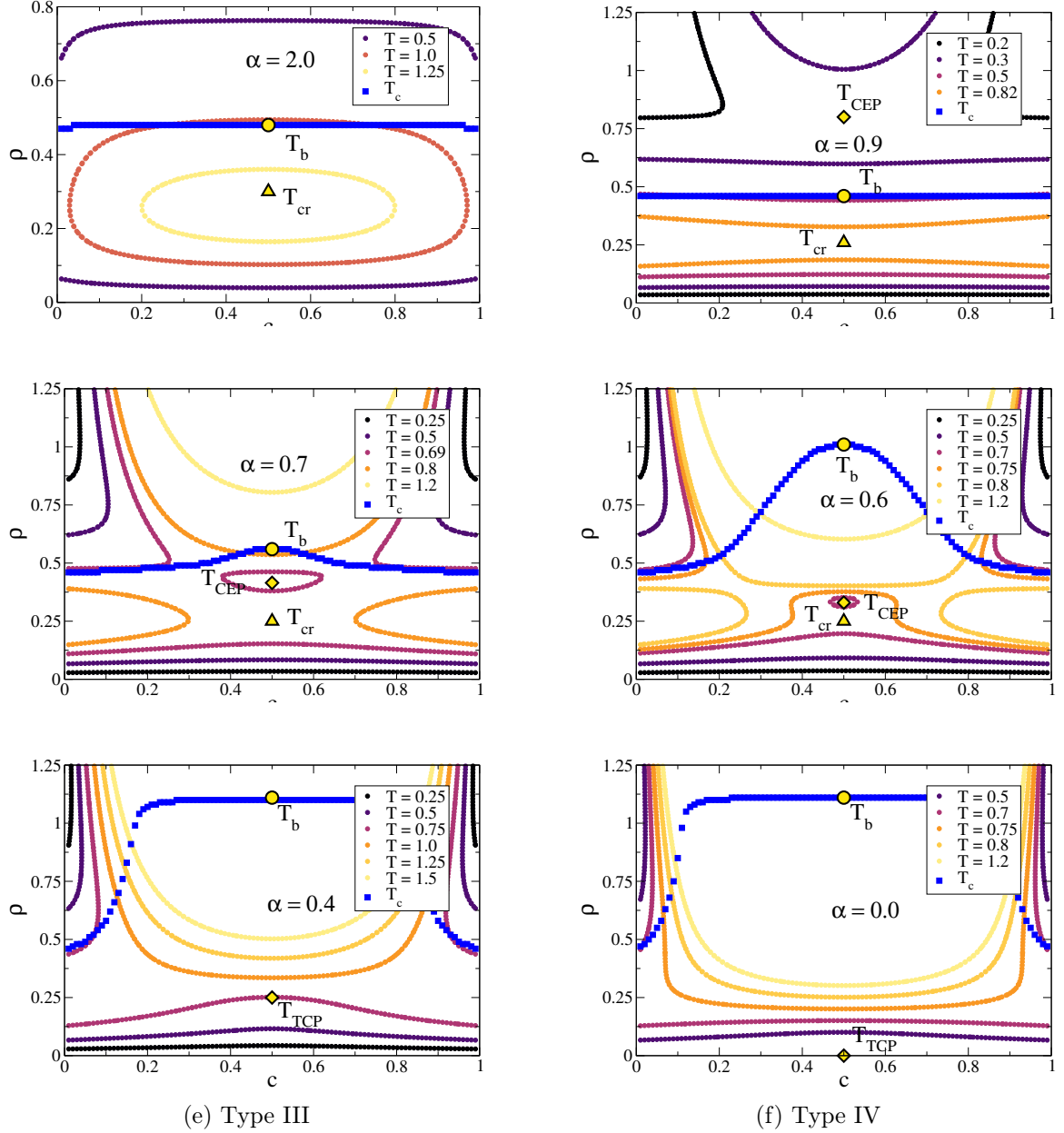

    \centering
    \begin{subfigure}[b]{0.48\textwidth}
        \centering
        \includegraphics[width=0.9\linewidth]{Fig_S6/alpha_2.0.eps}
        \caption{Case $\alpha = 2.0$}
    \end{subfigure}
    \hfill
    \begin{subfigure}[b]{0.48\textwidth}
        \centering
        \includegraphics[width=0.9\linewidth]{Fig_S6/alpha_0.9.eps}
        \caption{Type I}
    \end{subfigure}
    
    \begin{subfigure}[b]{0.48\textwidth}
        \centering
        \includegraphics[width=0.9\linewidth]{Fig_S6/alpha_0.7.eps}
        \caption{Type II-B}
    \end{subfigure}
    \hfill
    \begin{subfigure}[b]{0.48\textwidth}
        \centering
        \includegraphics[width=0.9\linewidth]{Fig_S6/alpha_0.6.eps}
        \caption{Type II-A}
    \end{subfigure}
    
    \begin{subfigure}[b]{0.48\textwidth}
        \centering
        \includegraphics[width=0.9\linewidth]{Fig_S6/alpha_0.4.eps}
        \caption{Type III}
    \end{subfigure}
    \hfill
    \begin{subfigure}[b]{0.48\textwidth}
        \centering
        \includegraphics[width=0.9\linewidth]{Fig_S6/alpha_0.0.eps}
        \caption{Type IV}
    \end{subfigure}
    
    \caption{
    Evolution of the Miscibility Gap.
    Iso-temperature contours of the instability surface $\lambda_1(c, \rho, T) = 0$ projected onto the $(\rho, c)$ plane for varying interaction strengths $\alpha$.
    Lighter lines indicate higher temperatures.
    (a-b) In the condensation-dominated regimes, contours form closed loops characteristic of simple fluids.
    (c-d) The emergence of demixing (Type II) distorts the high-density contours.
    (f) In the Type IV limit, contours become vertical lines, indicating a pure demixing instability driven solely by concentration.
    Solid blue squares marks the intersection of the instability surface $T_\text{inst}$ with the dynamic arrest surface $T_c$, locating the thermodynamic state points where the glass transition intervenes in the phase separation process.}
    \label{fig:S6}
\end{figure}

Figure \ref{fig:S6} presents complementary 2D projections of the instability surface $T_\text{inst}(\rho, c)$ onto the concentration-density plane for selected isotherms. 
These contour maps visualize the evolution of the \textit{miscibility gap} across the topological regimes:

\begin{itemize}
    \item \textbf{Condensation Dominance ($\alpha \ge 0.9$):} For Type I-A ($\alpha=2.0$) and Type I ($\alpha=0.9$), the isotherms form closed, concentric loops centered around the critical density. This topology is characteristic of gas-liquid condensation, where the instability is bounded in both density and concentration.
    
    \item \textbf{Emergence of Demixing ($\alpha \in [0.7, 0.4]$):} As the system traverses the Type II ($\alpha=0.7, 0.6$) and Type III ($\alpha=0.4$) regimes, the contours at high densities begin to flatten and widen. This distortion signifies the emergence of the liquid-liquid demixing instability ($S_{cc}$), which competes with and eventually disrupts the condensation dome~\cite{Koefinger2006, JAMR_2005}.
    
    \item \textbf{Pure Demixing ($\alpha = 0.0$):} In the Type IV limit, the contours transform into vertical lines (parallel to the density axis). This indicates that the instability condition becomes independent of total density and is driven exclusively by concentration fluctuations, consistent with the behavior of non-interacting species where mixing is purely entropic~\cite{Schoell2005}.
\end{itemize}

The solid blue squares mark the specific state points $(\rho, c)$ where the thermodynamic instability surface intersects the dynamic arrest surface $T_c$. These points define the \textit{kinetic locus} of the system: they represent the highest temperature at which the system encounters thermodynamic instability before arresting into a glass.

Finally, here is Table \ref{tab:kinetic_atlas} summarizing the main relations between the classification topology by Konynenburg and Scott~\cite{Konynenburg1980, Schoell2005, Koefinger2006}, and the kinetic perspective by the SCGLE theory\cite{Rigo2008, Rigo2008_2}.

\begin{table}[H]
    \centering
    \caption{Proposed Kinetic Atlas Classification. This table aligns the interaction parameter $\alpha$ used in this work with the standard thermodynamic topological classifications of van Konynenburg and Scott \cite{Konynenburg1980} and Köfinger \cite{Koefinger2006}. The ``Physical Name'' column establishes the terminology used to describe the specific non-equilibrium phenomena observed in our NESCGLE results.}
    \label{tab:kinetic_atlas}
    \renewcommand{\arraystretch}{1.3} 
    \begin{tabular}{c c >{\raggedright\arraybackslash}p{2.8cm} p{7cm}}
        \hline \hline
        \textbf{Energy Ratio} & \textbf{Literature} & \textbf{Physical Name} & \textbf{Key Topological \& Kinetic Feature} \\
        ($\alpha = \epsilon_{ij}/\epsilon_{ii}$) & \textbf{Type} & \textbf{(This Work)} &  \\
        \hline
        $\alpha > 1.0$ & \textbf{Type I-A} & \textbf{Pure Condensation} & 
        \textit{Demixing is thermodynamically impossible.} Cross-attraction dominates; the demixing spinodal solution vanishes \cite{Fantoni2005}. Arrest is driven strictly by density fluctuations ($S_{\rho\rho}$). \\
        
        $\alpha \approx 0.8$ & \textbf{Type I} & \textbf{Demixing frustration} & 
        \textit{Demixing exists but is buried.} The Critical End Point (CEP) lies deep within the spinodal. The glass transition line $T_c$ envelopes the instability, kinetically suppressing phase separation. \\
        
        $\alpha \approx 0.6$ & \textbf{Type II} & \textbf{Emergent Demixing} & 
        \textit{Competition regime.} The demixing instability ($T_\lambda$) rises in temperature. $T_c$ intersects the instability surface near the CEP \cite{JAMR_2005}, creating a window for gelation driven by phase separation. \\
        
        $\alpha \approx 0.3$ & \textbf{Type III} & \textbf{Tricritical Merger} & 
        \textit{Coalescence.} The Gas-Liquid and Liquid-Liquid instability lines merge at a Tricritical Point (TCP). Arrest is driven by a hybrid of density and concentration fluctuations. \\
        
        $\alpha \approx 0.0$ & \textbf{Type IV} & \textbf{Pure Demixing} & 
        \textit{Condensation vanishes.} In the absence of cross-attraction, the GL dome disappears. Instability is driven exclusively by Liquid-Liquid demixing ($S_{cc}$), bounded by hard-sphere glass dynamics. \\
        \hline \hline
    \end{tabular}
\end{table}

\section{The ``Gel-Glass'' Transition Line}

In this section, we present the results of a systematic evaluation of the asymptotic localization length (defined in eq. \eqref{eq:gammas}), $\gamma^{(a)}_A=\gamma^{(a)}_B=\gamma^{(a)}$, for a set of independent isochoric thermal quenches. 
Each quench proceeds from a common initial temperature $T_i = 10.0$ (well within the fluid region) to a variable final temperature $T$.

\begin{figure}[h]
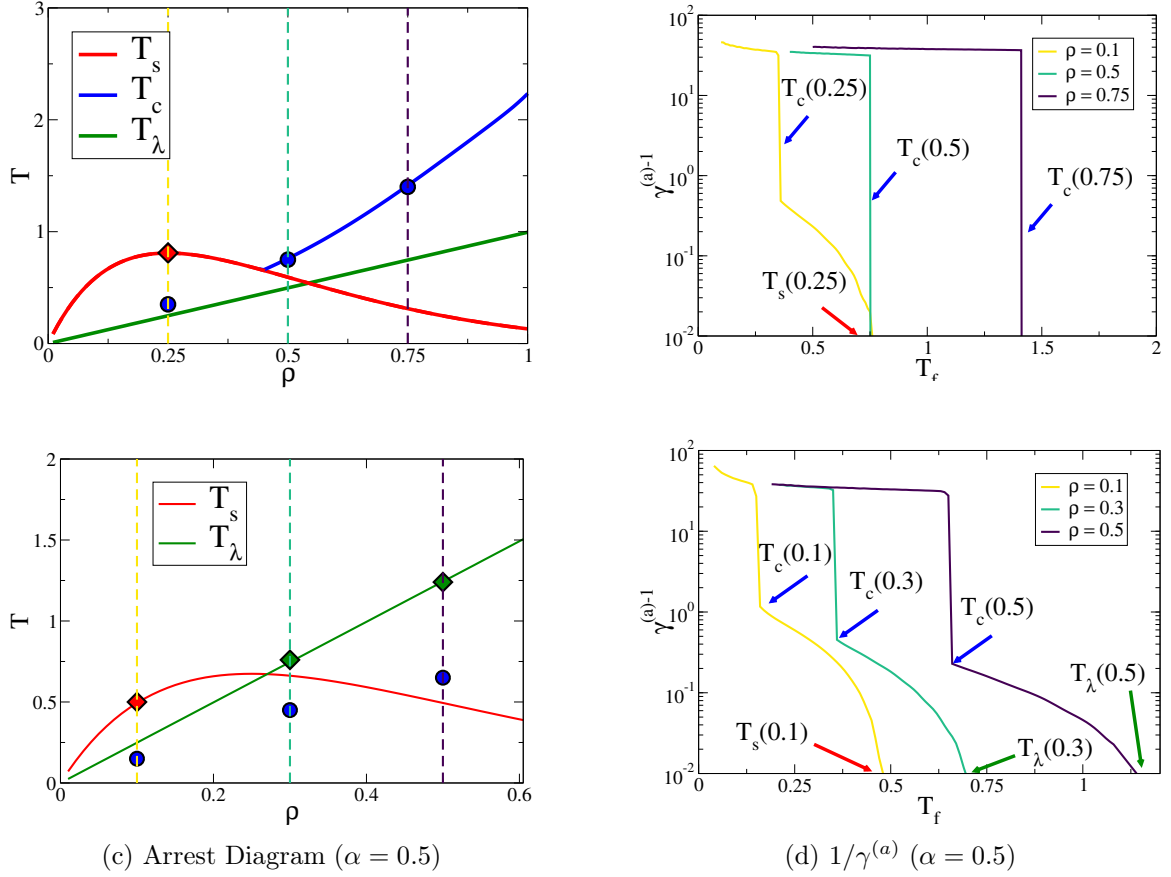

    \centering
    \begin{subfigure}[b]{0.48\textwidth}
        \centering
        \includegraphics[width=0.9\linewidth]{Fig_S7/alpha_0.8_c_0.5_SM.eps}
        \caption{Arrest Diagram ($\alpha = 0.8$)}
    \end{subfigure}
    \hfill
    \begin{subfigure}[b]{0.48\textwidth}
        \centering
        \includegraphics[width=0.9\linewidth]{Fig_S7/gammas_Uab_0.8_c_0.5.eps}
        \caption{$1/\gamma^{(a)}$ ($\alpha=0.8$)}
    \end{subfigure}
    
    \begin{subfigure}[b]{0.48\textwidth}
        \centering
        \includegraphics[width=0.9\linewidth]{Fig_S7/alpha_0.5_c_0.5_SM.eps}
        \caption{Arrest Diagram ($\alpha = 0.5$)}
    \end{subfigure}
    \hfill
    \begin{subfigure}[b]{0.48\textwidth}
        \centering
        \includegraphics[width=0.9\linewidth]{Fig_S7/gammas_Uab_0.5_c_0.5.eps}
        \caption{$1/\gamma^{(a)}$ ($\alpha=0.5$)}
    \end{subfigure}
    \caption{
    (a) Equilibrium SCGLE arrest diagram for the SBM ($c=0.5, \alpha = 2.0$). Dashed red line: Spinodal $T_s$. Solid blue line: Critical arrest temperature $T_c$.
    (b) Inverse asymptotic localization length $1/\gamma^{(a)}$ vs. temperature for selected isochores at $\alpha=2.0$.
    (c) Arrest diagram for $\alpha = 0.5$. 
    Dashed green line: Demixing $\lambda$-line $T_\lambda$.
    (d) Inverse asymptotic localization length $1/\gamma^{(a)}$ vs. temperature for selected isochores at $\alpha=0.5$.}
    \label{fig:NE_gammas}
\end{figure}

Figure \ref{fig:NE_gammas} displays the dynamic arrest diagrams and the corresponding inverse localization lengths for two distinct interaction regimes: the strong cross-attraction case ($\alpha = 2.0$, condensation-dominated) and the competitive interaction case ($\alpha = 0.5$, demixing rise).

\subsection{Case $\alpha = 0.8$}

The arrest diagram in Fig. \ref{fig:NE_gammas}(a) shows the spinodal temperature $T_s$ (solid red line), the $\lambda$-line $T_\lambda$ (solid green line), and the critical arrest temperature $T_c$ (solid blue line).

Figure \ref{fig:NE_gammas}(b) details the behavior of the inverse asymptotic localization length, $1/\gamma^{(a)}$, as a function of the final quench temperature $T$ for three representative isochores.
For high-density isochores ($\rho = 0.5$ and $0.75$) that do not intersect the spinodal region ($\rho > \rho_b$), the system undergoes a direct transition from an ergodic fluid to a non-ergodic glass.
As the temperature $T$ decreases, $1/\gamma^{(a)}$ jumps discontinuously from zero to a finite value on the order of $10^2$, corresponding to a localization length of $\sqrt{\gamma^{(a)}} \sim 0.1\sigma$.
This discontinuity marks the \textit{ideal glass transition} at $T_c$, where particle motion changes abruptly from diffusive exploration of the entire volume to localization within a cage approximately 10\% of the particle diameter~\cite{nescgle3}.

In contrast, the low-density isochore ($\rho = 0.25$) intersects the spinodal line $T_s$, resulting in a fundamentally different arrest scenario.
As the temperature is lowered towards $T_s$, the transition from the ergodic fluid is \textit{continuous}, characterized by $1/\gamma^{(a)}$ rising smoothly from zero, which indicates a critical slowing down associated with the thermodynamic instability.
However, at a lower temperature $T_c < T_s$, a second, discontinuous jump in $1/\gamma^{(a)}$ is observed.
In the intermediate regime $T_c < T < T_s$, the inverse localization length is of the order $10^{-2}$ (corresponding to $\sqrt{\gamma^{(a)}} \sim 10\sigma$), implying that particles are confined to mesoscopic domains rather than the tight cages of a glass.
This signature is characteristic of gel formation or arrested spinodal decomposition~\cite{nescgle5}.

\subsection{Case $\alpha = 0.5$}

Figure \ref{fig:NE_gammas}(c) presents the arrest diagram for $\alpha = 0.5$. 
Here, a second instability line appears: the $\lambda$-line (dashed green), indicating the onset of demixing. 
The dynamic arrest line $T_c$ intersects this $\lambda$-line at a critical density $\rho \approx 1.1$.

Figure \ref{fig:NE_gammas}(d) shows the corresponding localization lengths for the isochores $\rho = 0.1$, $0.3$, and $0.5$.
For the intermediate and high densities ($\rho = 0.3$ and $0.5$), the isochores intersect the $\lambda$-line before reaching the arrest line.
Similar to the spinodal crossing observed in the $\alpha=0.8$ case, we observe a two-step arrest process.
First, a continuous transition at $T_\lambda$ marks the onset of demixing kinetics, where $1/\gamma^{(a)}$ becomes finite but small ($\sim 10^{-1}$).
Subsequently, at a lower temperature $T_c < T_\lambda$, a discontinuous jump occurs as the dense, demixed phase undergoes structural arrest.
In contrast, the low-density isochore ($\rho = 0.1$) intersects the Gas-Liquid spinodal $T_s$.
The behavior here mirrors the $\alpha=0.8$ case, characterized by a continuous transition at $T_s$ followed by a discontinuous glass transition at $T_c$.

By systematically mapping these transitions across all isochores and interaction parameters $\alpha$, we identify that the lower critical temperature $T_c(\rho)$ (where the discontinuous jump occurs inside the instability region) is the continuation of the liquid-glass transition line $T_c(\rho)$ from the stable fluid region. 
This line defines a \textit{``gel-glass'' transition} separating two distinct non-ergodic states: the mesoscopically arrested gel/spinodal state (bounded by $T_s$ or $T_\lambda$) and the microscopically arrested attractive glass (bounded by $T_c$).

\section{Structural Overview} \label{sec:Results_Diagnostics}

In this supplemental, we provide a more detailed characterization than the main text, focused on the representation of fluctuation correlations.

\subsection{Quenches into the Condensation Region}

An additional characteristic that can be studied using the asymptotic material time $u^{(a)}$ is the asymptotic structure factor, defined as $\bm{S}^{(a)}(k) = \bm{S}(k;u^{(a)})$. 
The elements of this matrix, given by $S_{ij}^{(a)}(k) = \lim_{t\to\infty}(1/N)\langle\delta \rho_i(k;t)\delta \rho_j(-k;t)\rangle^{ss}$, represent the long-time correlations of the density fluctuations between species $i$ and $j$.

Following the Chen-Forstmann formalism \cite{Forstmann}, we adapt their representation to evaluate the asymptotic structure factors in the number-concentration basis. 
The transformed correlations take the form:
\begin{align}
    S^{(a)}_{\rho\rho}(k) &= c_A S^{(a)}_{AA}(k) + c_B S^{(a)}_{BB}(k) + 2\sqrt{c_A c_B} S^{(a)}_{AB}(k), \label{eq:S_rho_rho} \\
    S^{(a)}_{cc}(k) &= c_B S^{(a)}_{AA}(k) + c_A S^{(a)}_{BB}(k) - 2\sqrt{c_A c_B} S^{(a)}_{AB}(k), \label{eq:S_cc_A} \\
    S^{(a)}_{\rho c}(k) &= \sqrt{c_Ac_B} \left( S^{(a)}_{AA}(k) - S^{(a)}_{BB}(k) \right) - (c_A - c_B) S^{(a)}_{AB}(k). \label{eq:S_rho_c_A}
\end{align}

\begin{figure}
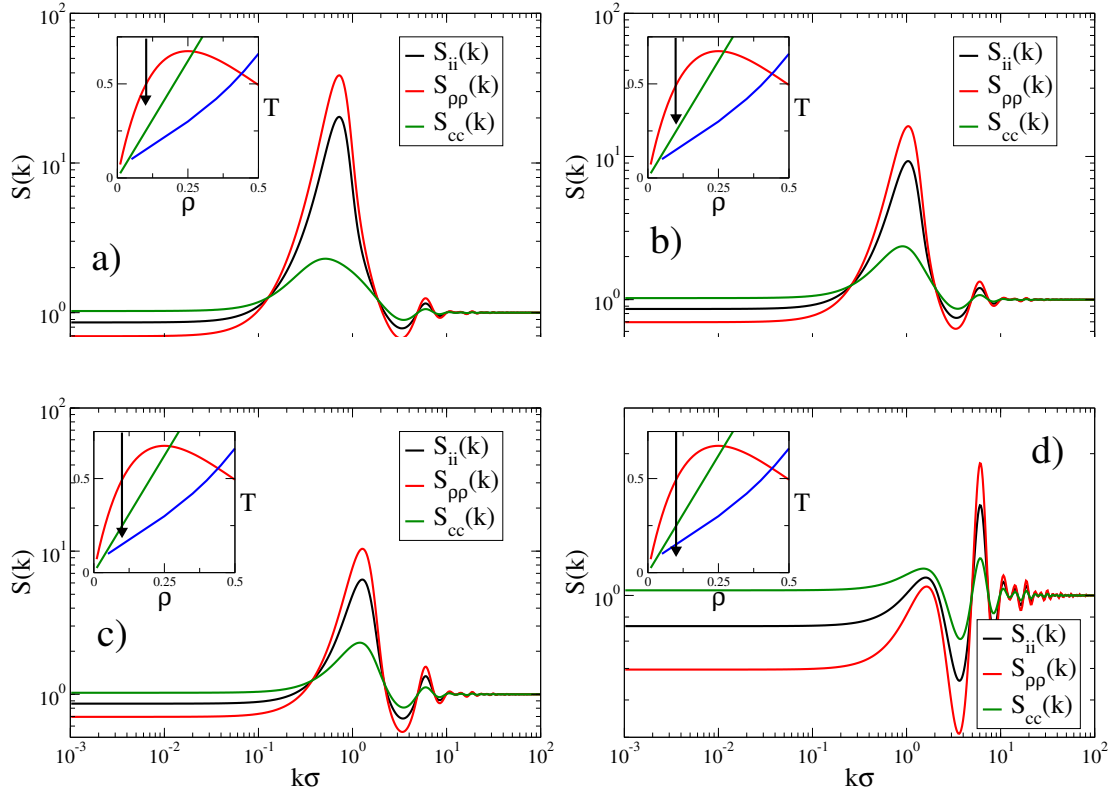

    \centering
    \includegraphics[width=0.45\linewidth]{Fig_S8/rho_0.1_Tf_0.4.eps}
    \includegraphics[width=0.45\linewidth]{Fig_S8/rho_0.1_Tf_0.3.eps}
    \includegraphics[width=0.45\linewidth]{Fig_S8/rho_0.1_Tf_0.2.eps}
    \includegraphics[width=0.45\linewidth]{Fig_S8/rho_0.1_Tf_0.1.eps}
    \caption{
    Asymptotic structure factors for the isochore $\rho=0.1$ in two representations: 
    $S_{ii}(k)\equiv (1/N)\angles{\delta\rho_i(k)\delta\rho_i(-k)}$ (solid black lines); 
    $S_{\rho\rho}(k)\equiv(1/N)\angles{\delta\rho(k)\delta\rho(-k)}$ (solid red lines), and $S_{cc}(k)\equiv(1/N)\angles{\delta c(k)\delta c(-k)}$ (solid green lines).
    The final temperature of each independent quench is (a) $T=0.4$, (b) $0.3$, (c) $0.2$, and (d) $0.1$.
    (Inset) Non-Equilibrium Arrest diagram from Figure 1(b) if the main text, with the final temperature indicated by a solid green triangle.}
    \label{fig:SA_rho_0.1}
\end{figure}

In Figure \ref{fig:SA_rho_0.1}, we examine the asymptotic structure factor using both the partial species representation ($S^{(a)}_{ii}$) and the number-concentration representation ($S^{(a)}_{\rho\rho}$ and $S^{(a)}_{cc}$) for the cross-interaction parameter $\alpha = 0.5$, equimolar concentration $c=0.5$, and the isochore $\rho=0.1$. 
This density allows us to explore the system's behavior within the condensation instability (Region III). 
We present results for four independent quenches to final temperatures $T_f=0.4, 0.3, 0.2$, and $0.1$.

Figure \ref{fig:SA_rho_0.1}(a) specifically illustrates the effect of a shallow quench to $T_f=0.4$, slightly below the spinodal temperature ($T_s=0.48$). 
The partial structure factors $S^{(a)}_{ii}(k)$ (solid black lines) exhibit a prominent peak at low wave-vectors, $k_{IRO}\simeq 0.5$. 
This peak, indicative of Intermediate Range Order (IRO) structures, is significantly higher than the main structural peak at $k_{main}\simeq 2\pi$ (i.e., $S^{(a)}_{ii}(k_{IRO}) > S^{(a)}_{ii}(k_{main})$).
In the number-concentration representation, the total density-density correlation $S^{(a)}_{\rho\rho}(k)$ (solid red lines) follows the same trend, displaying a pronounced low-$k$ peak. 
The concentration-concentration correlation $S^{(a)}_{cc}(k)$ (solid green curves) also exhibits this feature, but with a much lower intensity; notably, $S^{(a)}_{\rho\rho}(k_{IRO}) \gg S^{(a)}_{cc}(k_{IRO})$. 
Finally, for the equimolar concentration ($c=0.5$), the cross-correlation term vanishes identically ($S^{(a)}_{\rho c}(k) = 0$ for all $k$) due to symmetry [Eq. \eqref{eq:S_rho_c_A}].

As we lower the temperature, the magnitude of $S^{(a)}_{\rho\rho}(k_{IRO})$ decreases as the final temperature approaches the $\lambda$-line at $T_\lambda=0.29$. 
Figures \ref{fig:SA_rho_0.1}(b) and (c) illustrate the system's behavior as it crosses this threshold (for $T_f = 0.3$ and $0.2$, respectively); 
remarkably, we observe no qualitative influence of the transition temperature on the structural evolution. 
The most notable characteristic of this cooling process is the significant suppression of the IRO peak height, which drops by nearly an order of magnitude. 
In contrast, crossing the dynamic arrest line ($T_c = 0.12$) has a profound impact on the asymptotic structure. 
As shown in Fig. \ref{fig:SA_rho_0.1}(d) for $T_f=0.1 < T_c$, the IRO peak is suppressed to such an extent that the main structural peak becomes the dominant feature.

Throughout this regime, the number-number structure factor, $S^{(a)}_{\rho\rho}$, effectively captures the dominant behavior observed in the partial structure factors, $S^{(a)}_{ii}$. 
This consistency reflects the physical nature of the spinodal line $T_s(\rho)$, which arises specifically from instabilities in density correlations. 

\subsection{Quenches into the Demixing Region}

\begin{figure}
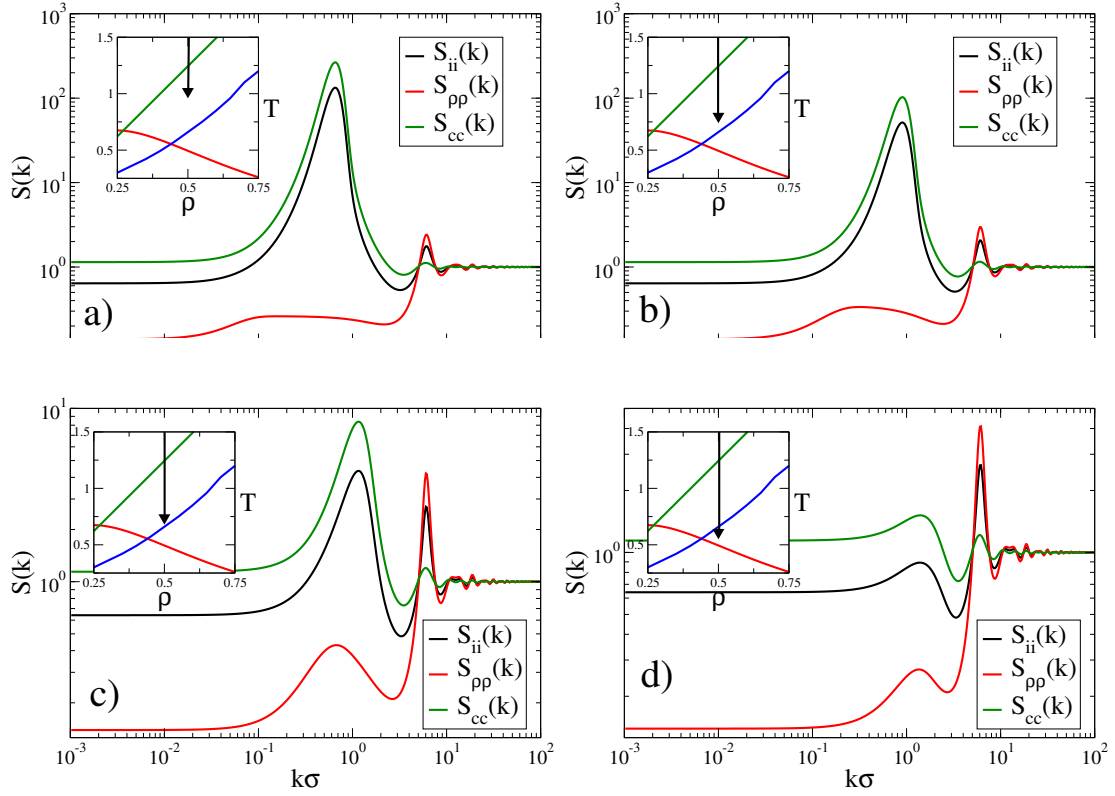

    \centering
    \includegraphics[width=0.45\linewidth]{Fig_S9/rho_0.5_Tf_1.0.eps}
    \includegraphics[width=0.45\linewidth]{Fig_S9/rho_0.5_Tf_0.8.eps}
    \includegraphics[width=0.45\linewidth]{Fig_S9/rho_0.5_Tf_0.64.eps}
    \includegraphics[width=0.45\linewidth]{Fig_S9/rho_0.5_Tf_0.6.eps}
    \caption{
    Asymptotic structure factors for the isochore $\rho=0.5$ in two representations: 
    The $(\rho_1,\rho_2)$-representation $S_{ii}(k)$ (solid black lines); 
    and the $(c, \rho)$-representation $S_{\rho\rho}(k)$ (solid red lines) and $S_{cc}(k)$ (solid green lines).
    The final temperature of each independent quench is (a) $T=1.0$, (b) $0.8$, (c) $0.64$, and (d) $0.6$.
    (Inset) Non-Equilibrium Arrest diagram from Figure 1(b) if the main text, with the final temperature indicated by a solid green triangle.}
    \label{fig:SA_rho_0.5}
\end{figure}

Next, we explore the behavior within the demixing regime (Region IV). 
In Figure \ref{fig:SA_rho_0.5}, we plot the asymptotic structure factor for the isochore $\rho=0.5$. 
First, Figure \ref{fig:SA_rho_0.5}(a) shows the result for a final temperature of $T_f=1.0$, which lies below the $\lambda$-line ($T_\lambda=1.45$). 
We then progressively decrease the temperature to $T_f = 0.8$ and $0.64$ (Figs. \ref{fig:SA_rho_0.5}(b) and (c), respectively), remaining above the dynamic arrest temperature $T_c=0.62$. 
Finally, in Figure \ref{fig:SA_rho_0.5}(d), we present the case where the temperature is lowered to $T_f = 0.6$, crossing into the arrested regime ($T_f < T_c$).

Focusing first on the partial structure factors $S^{(a)}_{ii}$ (solid black lines), we observe only the formation of IRO structures associated with the low-$k$ peak, which is notably suppressed below the transition line $T_c(\rho)$. 
However, the number-concentration representation provides a much richer perspective, revealing a complete inversion of the roles of the two fluctuation modes compared to the previous case. 
Here, in the demixing Region IV, the concentration-concentration structure factor $S^{(a)}_{cc}$ (solid green line) dominates the behavior, reflecting the strong concentration fluctuations characteristic of demixing. 
Crucially, below the arrest temperature $T_c$, the main structural peak is almost entirely accounted for by the density correlations $S^{(a)}_{\rho\rho}$ (solid red line). 
This behavior effectively decouples the nature of the two peaks: the low-$k$ IRO peak is driven by concentration fluctuations (demixing), while the high-$k$ main peak is driven by density fluctuations (local packing/arrest).

Note that the partial structure factors in Fig. \ref{fig:SA_rho_0.1} (condensation) and Fig. \ref{fig:SA_rho_0.5} (demixing) exhibit nearly identical low-k features, confirming the structural blindness hypothesis
The scenarios depicted in Figures \ref{fig:SA_rho_0.1} (predominantly condensation) and \ref{fig:SA_rho_0.5} (predominantly demixing) reveal the behavior of the static structure factor within the instability regions III and IV. 
Both regimes share common features: the formation of an IRO peak at low $k$-values, and its sudden suppression when the final temperature enters the Double Glass region (Region II). 
Crucially, the partial structure factor representation, $S^{(a)}_{ii}$, is incapable of differentiating the nature of the increased correlations in these two regimes. 
In contrast, the number-concentration representation provides a clear distinction. 
In the condensation region, the IRO peak is fully explained by the total density fluctuations ($S^{(a)}_{\rho\rho}$), whereas in the demixing region, the concentration fluctuations ($S^{(a)}_{cc}$) are sufficient to account for the IRO peak.

\bibliographystyle{unsrt}
\bibliography{biblio_SM}